\documentclass[letter]{aa}

\usepackage{graphicx}
\usepackage{txfonts}
\usepackage[colorlinks = true,
           linkcolor = blue,
            urlcolor  = blue,
            citecolor = blue,
            anchorcolor = blue]{hyperref}
\begin{document}

   \title{NOEMA reveals the true nature of luminous red JWST $z>10$ galaxy candidates}

   \author{R. A. Meyer
          \inst{1}\fnmsep \inst{2} \orcid{0000-0001-5492-4522}
          \and
          L. Barrufet \inst{1} \orcid{0000-0003-1641-6185}
          \and L. A. Boogaard \inst{2} \orcid{0000-0002-3952-8588}
          \and R. P. Naidu \inst{3}\fnmsep \inst{4} \orcid{0000-0003-3997-5705}
          \and P. A. Oesch \inst{1,5} \orcid{0000-0001-5851-6649}
          \and F. Walter \inst{2} \orcid{0000-0003-4793-7880}
          }

   \institute{Departement d'Astronomie, University of Geneva, Chemin Pegasi 51, 1290 Versoix, Switzerland\\
              \email{romain.meyer@unige.ch}
         \and
             Max-Planck Institute for Astronomy, K\"onigstuhl 17, 69118 Heidelberg, Germany
         \and
            Center for Astrophysics | Harvard \& Smithsonian, 60 Garden Street, Cambridge, MA 02138, USA
                \and    
            MIT Kavli Institute for Astrophysics and Space Research, 77 Massachusetts Ave., Cambridge, MA 02139, USA
        \and
            Cosmic Dawn Center (DAWN), Niels Bohr Institute, University of Copenhagen, Jagtvej 128, K\o benhavn N, DK-2200, Denmark\\     
             }

   \date{Received October --, 2023; accepted --}

% \abstract{}{}{}{}{}
% 5 {} token are mandatory
  \abstract{
   The first year of JWST has revealed a surprisingly large number of luminous galaxy candidates beyond $z>10$. While some galaxies have already been spectroscopically confirmed, there is mounting evidence that a subsample of the candidates with particularly red inferred UV colours are, in fact, lower redshift contaminants.
    These interlopers are often found to be `HST-dark' or `optically faint' galaxies at $z\sim2-6$, a population that is key to improving our understanding of dust-obscured star formation throughout cosmic time.
   This paper demonstrates the complementarity of ground-based mm-interferometry and JWST infrared imaging to unveil the true nature of red  1.5-2.0 $\mu \rm{m}$ dropouts that have been selected as ultra-high-redshift galaxy candidates.
   We present NOEMA Polyfix follow-up observations of four JWST red 1.5-2.0 $\mu \rm{m}$ dropouts selected by \citet{Yan2023_darkfield} as ultra-high-redshift candidates in the PEARLS-IDF field. The new NOEMA observations constrain the rest-frame far-infrared continuum emission and efficiently discriminate between intermediate- and high-redshift solutions.
   We report $>10\sigma$ NOEMA continuum detections of all our target galaxies at observed frequencies of $\nu = 236$ and $252\ \rm{GHz}$, with FIR slopes indicating a redshift of $z<5$. We modelled their optical-to-FIR spectral energy distribution (SED) with multiple SED codes, finding that they are not $z>10$ galaxies but dust-obscured, massive star-forming galaxies at $z\sim 2-4$ instead. The contribution to the cosmic star formation rate density (CSFRD) of such sources is not negligible at $z\simeq 3.5$ ($\phi\gtrsim(1.9-4.4)\times10^{-3}\ \rm{cMpc}^{-3}$; or $>3-6\%$ of the total CSFRD), in line with previous studies of optically faint and sub-millimeter galaxies. This work showcases a new way to select intermediate- to high-redshift dust-obscured galaxies in JWST fields with minimal wavelength coverage. This approach opens up a new window onto obscured star formation at intermediate redshifts, whilst removing contaminants with red colours from searches at ultra-high redshifts.}

   \maketitle
%________________________________________________________________

\section{Introduction}

The first year of operation of JWST has transformed our view of early galaxy evolution. In particular, the imaging sensitivity and near-infrared coverage up to $\sim 5 \mu\rm{m}$ of the  NIRCam instrument \citep{nircam_performance} has been unveiling new galaxies undetected even in the deepest HST observations. These objects fall in two broad categories: 1) ultra-high-redshift candidates ($z>10$) invisible in HST due to the Lyman-$\alpha$ break caused by the neutral intergalactic medium and 2) a collection of massive, intermediate-redshift ($2\lesssim z \lesssim7$) galaxies with red colours and high dust attenuation making them extremely faint at the wavelengths probed by HST ($\lesssim 1.5 \mu \rm{m}$). 

Ultra-high-redshift galaxies have been the focus of multiple studies in the first deep fields observed with JWST, with numerous candidates claimed at $z>10$, as well as some up to $z>15$ \citep{Finkelstein2022_maisie,Finkelstein2023,Naidu2022,Harikane2023,Donnan2022,Adams2022a,Yan2023_darkfield,Yan2023_ero}. An excess at the bright end of the $z\gtrsim10$ UV luminosity function could indicate enhanced star formation rate efficiencies in the first 500 Myr, but caution is still necessary as most of the objects still only have photometric redshifts. Whilst multiple objects have been successfully confirmed with spectroscopy \citep[e.g.][]{ArrabalHaro2023, ArrabalHaro2023a,WangB2023, Harikane2023_specUVLF, Bunker2023_JADESrelease, Curtis-Lake2023}, catastrophic outliers have already been identified, with $z\sim5$ galaxies with extreme dust and/or line emission properties masquerading as $z>10$ candidates \citep{Naidu2022a,Zavala2023_z5dusty,ArrabalHaro2023,WangB2023}. Some studies, recognising the likely contamination of high-redshift samples by lower redshift dusty galaxies, have already proposed additional criteria (e.g. $\Delta \chi^2$ between low- and high-z solutions, integral of $p(z>7,8,9)$) to complement the initial photometric redshift selection \citep[e.g.][]{Finkelstein2023,Harikane2023}. These different definitions of robust high-redshift candidate lead to disagreements on the exact number of candidates in each field \citep{Bouwens2023a}. Whilst such ad hoc prescription have successfully led to large fraction of objects confirmed with spectroscopy \citep[e.g.][]{Finkelstein2023a,Harikane2023_specUVLF}, the nature of the dusty, likely lower-redshift contaminants remains speculative in the absence of dedicated follow-up observations.

The JWST-detected population of dusty $3\lesssim z\lesssim 6$ galaxies has been identified as an extension of `HST-dark', `ALMA-only', optically faint infrared, or dusty star-forming galaxies (DSFGs) thought to be responsible for most of the obscured star formation at high redshift \citep[e.g.][]{Blain1999,SMail2002,Chapman2005,Barger2012,Swinbank2014, Elbaz2011,Casey2014, Casey2018,TWang2019,Alcalde-Pampliega2019,Williams2019, Fudamoto2020,Dudzeviciute2020,Shu2022,Xiao2023}. Their detection and study with JWST shows that these names only described a particular selection function for objects with a wide range of properties \citep{Nelson2022, Rodighiero2022,Perez-Gonzalez2022,Barrufet2023, Smail2023,Barger2023}. Broadly speaking, these studies find that optically faint galaxies detected now with JWST are intermediate- to high-redshift ($2\lesssim z\lesssim 8$), dusty ($A_v>2$) with a large fraction of obscured star formation or passive galaxies, and could contribute a significant fraction $>10\%$ of the cosmic star formation rate density at $z\gtrsim5$. However, this picture is evolving rapidly as their census is still incomplete and new ways of identifying such optically invisible galaxy are still in development. In particular, large numbers of these objects might be hidden in ultra-high-redshift candidates lists, based on the detection of a break at $1-2\ \mu\rm{m}$.

The purpose of this Letter is to demonstrate the power of millimeter observations to determine the nature of ambiguous 1.5--2.0$\mu\rm{m}$ red dropouts detected in JWST imaging data. Specifically, we present follow-up observations of four  1.5--2.0$\ \mu\rm{m}$ red dropouts selected by \citet{Yan2023_darkfield} in the PEARLS survey with a limited number of JWST/NIRCam imaging filters (F150W, F277W, F356W, and F444W). We show that most SED fitting codes, using standard parameters, either prefer  $z>10$ or have degenerate or bimodel posterior redshift solutions. Here, we present  NOEMA 1 mm detections of their FIR continuum ($100\%$ detection rate) to demonstrate that such observations break the degeneracy between intermediate- and high-redshift photometric redshift solutions. Our SED modelling of the red 1.5--2.0$\mu\rm{m}$ dropouts shows that they are massive, dusty galaxies at $2<z<4,$ similar to so-called `HST-dark' galaxies \citep[see also][]{Zavala2023_z5dusty}.

Throughout this letter, we use a concordance cosmology with $H_0 = 70\ \rm{Mpc}\ \rm{km\ s}^{-1}$, $\Omega_M = 0.3$, and $\Omega_\Lambda = 0.7$. Magnitudes are indicated in the AB system \citep{Oke1974}.

\section{NOEMA observations}

We studied four objects detected in JWST F200W, F356W, and F444W imaging of the PEARLS survey \citep{Windhorst2023} and selected as ultra-high-redshift candidates based on their colours \citep{Yan2023_darkfield}. The four targets of this work were chosen to be the brightest, but also the reddest in the \cite{Yan2023_darkfield} sample (F150D\_JWIDF\_E01,F150D\_JWIDF\_H17, F200D\_JWIDF\_M03, and F200D\_JWIDF\_H08), guaranteeing a $\rm{S/N}\sim10$ detection at $\sim 250 \rm{GHz}$ if their redshift is $z\sim 3-5$, or $\rm{S/N}>\sim100$ if they are at $z>10$ (see further Section \ref{sec:SED} and Appendix \ref{appendix:A}). For completeness, their HST and JWST photometry from \cite{Yan2023_darkfield} is reproduced in Table \ref{table:1}, alongside our NOEMA FIR continuum measurements.

%-----------------------------------------------------------------
   \begin{figure*}[h!]
   \centering
    \includegraphics[height=0.195\textheight, trim=0 0 4cm 0cm,clip]{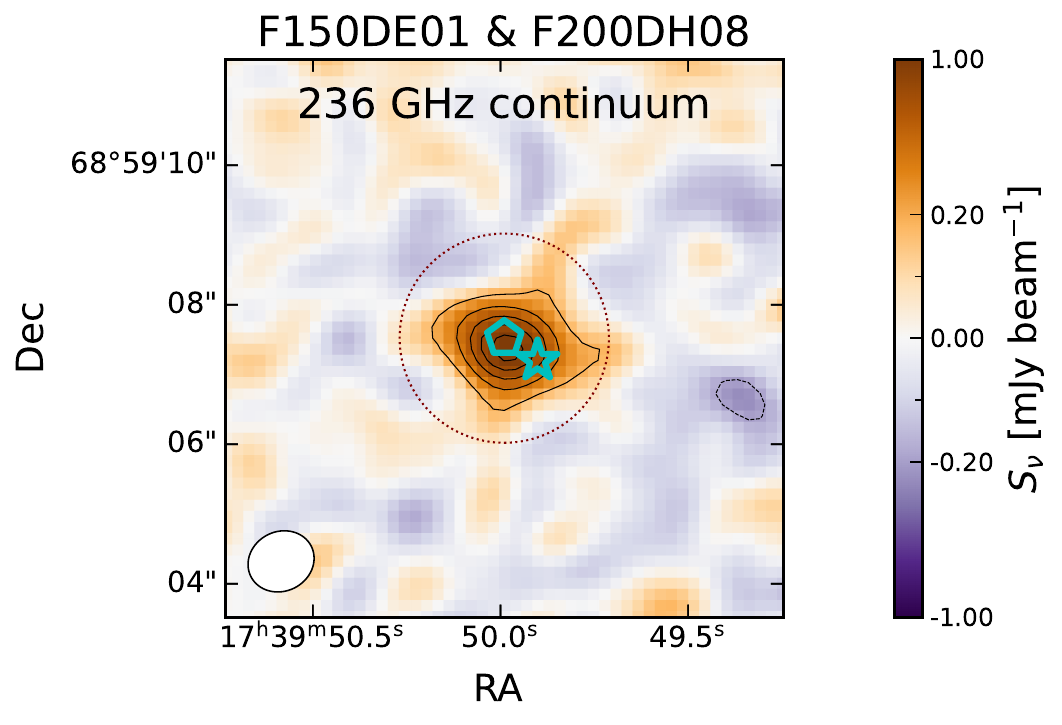}
    \includegraphics[height=0.195\textheight, trim=0 0 4cm 0cm,clip]{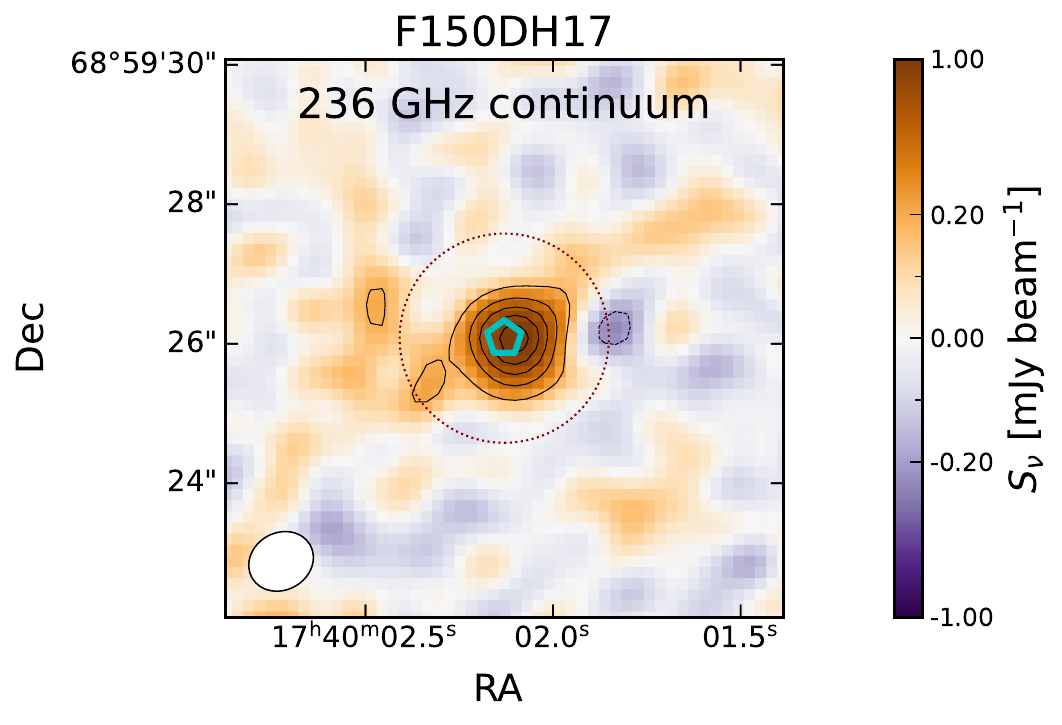}
    \includegraphics[height=0.195\textheight]{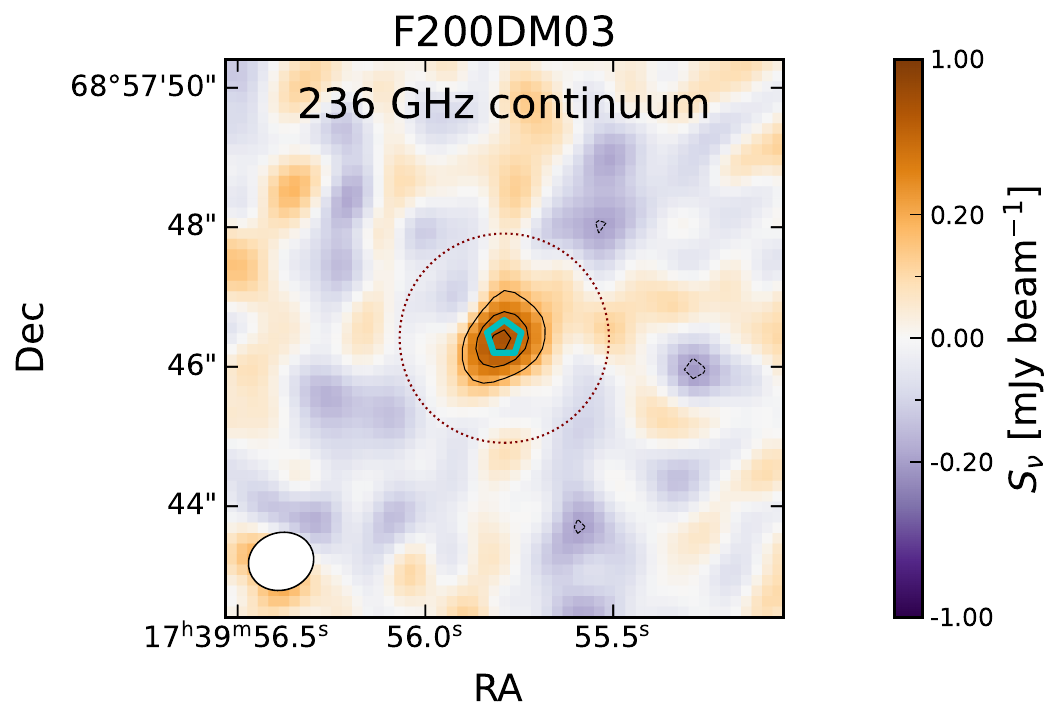} \\
    \includegraphics[height=0.182\textheight, trim=0 0 4cm 0.8cm,clip]{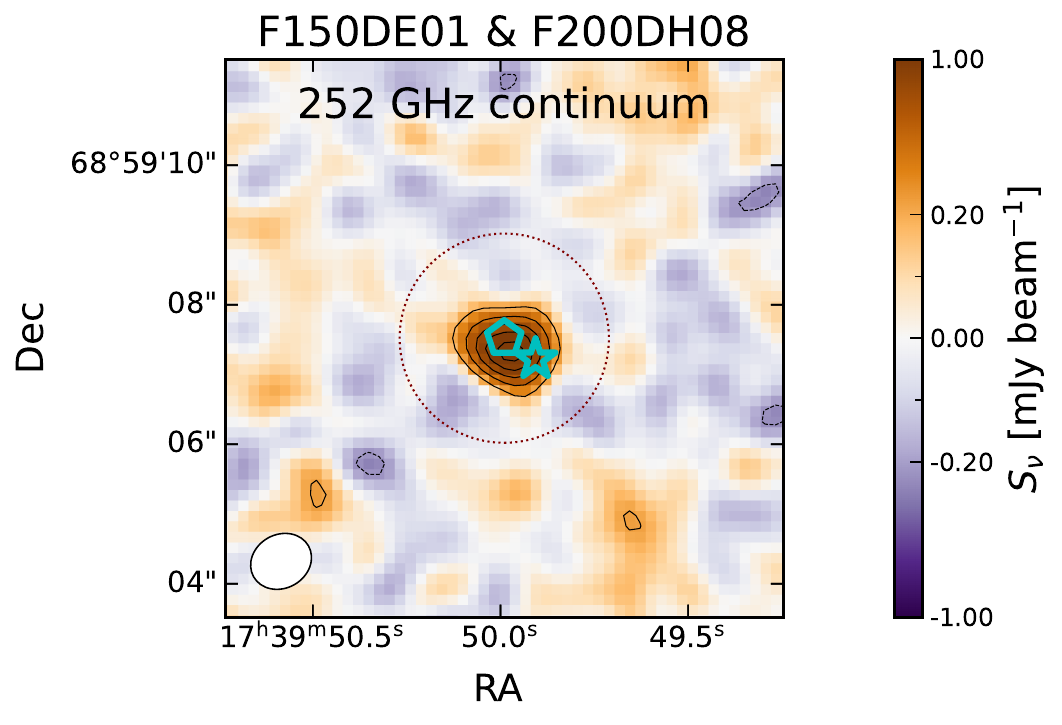}
    \includegraphics[height=0.182\textheight, trim=0 0 4cm 0.8cm,clip]{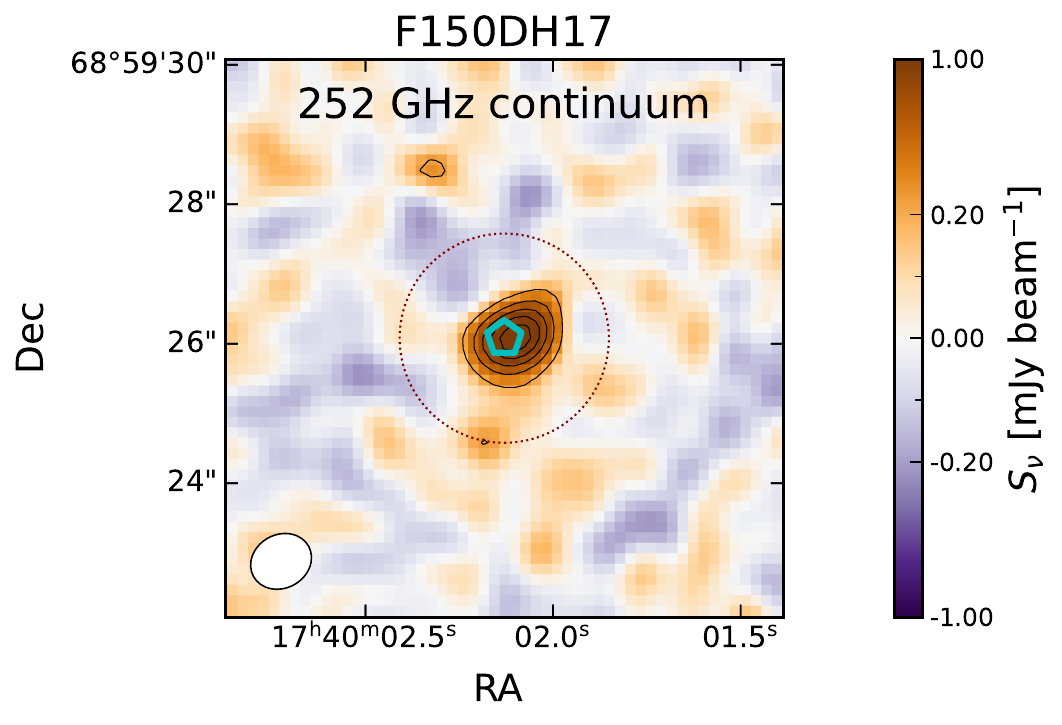}
    \includegraphics[height=0.182\textheight, trim=0 0 0 0.8cm,clip]{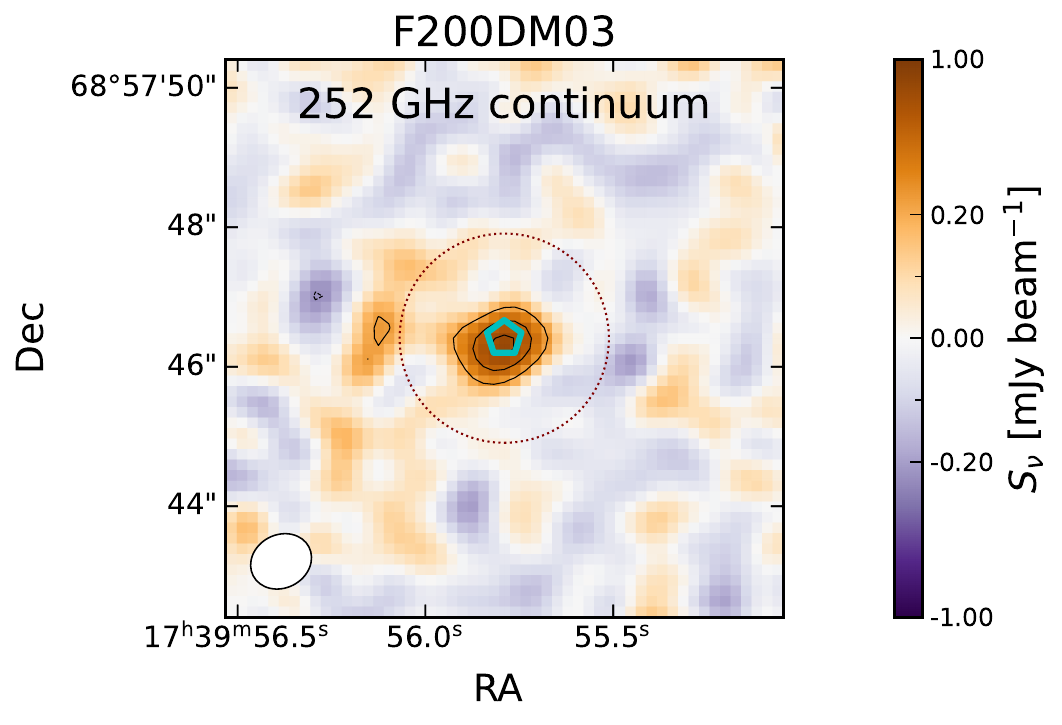} \\
      \caption{NOEMA continuum observations of the four red 1.5-2.0 $\mu\ \rm{m}$ dropouts from the PEARLS-IDF data in \citet[][indicated with cyan symbols]{Yan2023_darkfield}. We show the $236\ \rm{GHz}$ (top row) and $252\ \rm{GHz}$ continuum (bottom row) for each object with the galaxies position from JWST indicated cyan symbols. The full contours indicate the $(3,6,9,15) \sigma$ level, where $\sigma$ is the root mean square (rms) noise in the images. The dashed contours show the $-3\sigma$ level. The red dashed circle shows a $r=2\arcsec$ aperture.
      }
         \label{fig:noema_continuum}
   \end{figure*}
%-----------------------------------------------------------------
\renewcommand{\arraystretch}{1.5} 
\begin{table*}

\caption{Observed photometry of the PEARLS-IDF targets studied in this work. }             % title of Table
\label{table:1}      % is used to refer this table in the text
\centering      % used for centering table
\small
\begin{tabular}{l c c c c c c c}        % centered columns 
    \hline \hline% inserts double horizontal lines
ID & $m_{\rm{F814W}}$ [AB] & $m_{\rm{F150W}}$  [AB] & $m_{\rm{F200W}}$  [AB] & $m_{\rm{F356W}}$ [AB]  & $m_{\rm{F444W}}$  [AB] & $f_{252 \ \rm{GHz}}$ [mJy] & $f_{236 \ \rm{GHz}}$ [mJy]  \\    % table heading
\hline                        % inserts single horizontal line
F150D\_JWIDF\_E01 & $>28.47$ & $26.28\pm0.29$ & $25.07\pm0.08$ & $23.83\pm0.01$ & $23.27\pm0.01$ & $0.60\pm0.22$ & $0.52\pm0.18$  \\
F150D\_JWIDF\_H17  & $>28.47$ & $26.86\pm0.48$ & $25.64\pm0.14$ & $24.51\pm0.02$ & $24.01\pm0.01$ & $1.34\pm0.08$ & $1.08\pm0.07$  \\   % inserting body of the table
F200D\_JWIDF\_H08 & $>28.47$ & $>28.61$ & $28.23\pm0.35$ & $26.27\pm0.02$ & $25.53\pm0.01$ & $0.60\pm0.22$ & $0.52\pm0.18$ \\
F200D\_JWIDF\_M03   & $>28.47$ & $>27.45$ & $26.55\pm0.21$ & $24.05\pm0.01$ & $23.26\pm0.01$ &  $0.75\pm0.07$  & $0.62 \pm 0.06$ \\
\hline                                   %inserts single line
\end{tabular}
\tablefoot{The measured HST and JWST photometry is reproduced from \citet{Yan2023_darkfield} for completeness. Upper limits are given at the $2\sigma$ level.}
\end{table*}
%-----------------------------------------------------------------
The targets were observed with NOEMA in Band 3 between December 17-18 2022. The targets were set at the phase center and observed for $0.5$h (on-source) in track-sharing mode. Two targets are close (potentially interacting/merging; F150DE01 and F200DH08) and are observed simultaneously in the same pointing. The spectral setups were tuned at a nominal central frequency of $251.76\ \rm{GHz}$ in the upper sideband and $236.25\ \rm{GHz}$ in the lower sideband.

The data was calibrated and reduced at IRAM (remotely) using the latest \textit{CLIC} package in the GILDAS framework. We flag a small fraction of the tracks affected by poor weather conditions and baselines with bad phase solutions. We image the continuum in the lower and upper sidebands separately using natural weighting and Hogbom cleaning down to $2\sigma$ in \textit{MAPPING} with $r=5\arcsec$ circular support regions centered on the JWST target positions. The final synthesized beam size is $0.96\arcsec\times0.83\arcsec$ ($0.90\arcsec\times0.77\arcsec$) and the final continuum rms achieved is $62-64\ (73-76)\ \mu\rm{Jy\ beam}^{-1}$ at $236.25\ (251.76)\ \rm{GHz}$, respectively.

We show the continuum imaging of the targets in Fig. \ref{fig:noema_continuum}, where all the sources are clearly detected at $\rm{S/N}\simeq 10$--15. No additional continuum sources were detected in the NOEMA field of view ($\theta\sim 30\arcsec$). The sources are only marginally resolved and the difference between the aperture-integrated fluxes and that of the central pixel is at most $10-20\%,$ depending on the aperture size used. Thus, we  simply measured the continuum fluxes in the central brightest pixel, which is equivalent to the total flux within one (synthesised) beam. This also avoids any complications when comparing the fluxes between the two sidebands. 

The two close galaxies F150DE01 and F200DH08 are not resolved, with a total continuum flux of $f_{236\ \rm{GHz}} = 1.04 \pm 0.06 \ \rm{mJy}, f_{252\rm{GHz}} = 1.19 \pm 0.07\ \rm{mJy}$. Based on the assumption that both objects are of a similar nature, we assign half of the flux to each galaxy and use a fiducial relative error of $34\%$ reflecting the uncertainty on the fraction of flux originating from each object. This has little impact on our results as the redshift and nature of the sources is already sufficiently constrained by $\lesssim 10\ \rm{mJy}$ continuum detection. The continuum flux densities and errors for all sources  are presented in Table \ref{table:1}. 

We also produced datacubes for each sideband with 20 MHz-wide channels. The rms noise in the cube $1.2 (1.4)\ \rm{mJy\ beam}^{-1}$ per channel. For each source, we extract spectra in $r=2\arcsec$ apertures as well as in the central pixel only and we find no emission line in the frequency range covered by our observations ($232.5-240$ GHz, $248-255.5$ GHz).

%______________________________________________________________

\section{Multi-wavelength SED modelling and the nature of red 1.5--2.0\,$\mu\rm{m}$ dropouts}
\label{sec:SED}

We modelled the SED of the four red dropouts using five different codes: \emph{EAZY} \citet{Brammer2012}, \emph{CIGALE} \citep{Boquien2019}, \emph{BAGPIPES} \citep{Carnall2018}, and \emph{PROSPECTOR} \citet{Johnson2021}. We briefly review the parameters used to run each code below.  
\paragraph{EAZY: } The most important choice when using \textit{EAZY} is the choice of templates. We ran \emph{EAZY} twice: once with the \emph{tweak\_FSPS\_templates}, augmented with the three bluer templates (with and without Lyman-$\alpha$ emission) from \cite{Larson2022b}, and a second time with the \emph{blue\_sfhz}\footnote{https://github.com/gbrammer/eazy-photoz/tree/master/templates/sfhz} templates with redshift-dependent star formation histories. We fit the data between redshift 0 and 20, using no prior on the luminosity, and applied template error using the values derived from the COSMOS2020 dataset.
\paragraph{CIGALE: } We ran a grid of CIGALE models using a delayed star formation rate history ($0.1<\tau / [Gyr]<5$) and a recent burst, using the SSP models  from \citet{Bruzual2003} with a \citet{Chabrier2003} IMF and a fixed metallicity of $Z=0.02$. We varied the nebular line emission contribution with $-4<\log U<-1$ (in steps of 0.5) with a fixed gas metallicity of $Z_{\rm{gas}}=0.02$ and fixed electron density of $n_e=100\ cm^{-3}$. The dust attenuation was modelled with the \emph{dustatt\_modified\_starburst} module that uses the \citet{Calzetti2000} attenuation for the continuum, with $0<A_{V}<8.2$ and a MW-like attenuation curve for the emission lines. The models were fitted within $0.0<z<20$ in steps of $\Delta z = 0.05$. 

We also fit the near-infrared to FIR photometry with CIGALE templates, including an AGN contribution using SKIRTOR templates \citep[][]{Stalevski2016}. We find that the addition of AGN template does not improve the $\chi^2$ of the best-fit solutions and that the contribution of AGN to the observed photometry is negligible in the best-fit composite SEDs. We thus proceed by fitting the available photometry with pure stellar light SEDs.

\paragraph{BAGPIPES: } We used the latest version of \emph{BAGPIPES} \citep{Carnall2018} with a delayed star formation history ($0.1<\tau / [Gyr]<10$), the dust attenuation from \citet{Calzetti2000},  with $0<A_v<8$, nebular line emission with $-4< \log U <-2$, metallicities spanning $0-2.5\ Z_\odot$. \emph{BAGPIPES} uses \citet{Bruzual2003} SSP models with a \citet{Kroupa2001} IMF, and we use a fixed metallicity $Z=0.02$, matching that used in CIGALE. As \emph{BAGPIPES} does not include a special treatment of non-detections, we set the fluxes of upper limits to $0$ and use the $1\sigma$ upper limit as the error when fitting the spectra.

\paragraph{Prospector: } We ran \emph{Prospector} using a uniform redshift prior $0.1<z<20$ and a delayed-tau SFH history. The physical parameters priors follow the choices of \citet{Tacchella2022}, for instance, the gas and stellar metallicity, dust properties, and nebular emission parameters \citep[see  Table 1 and Section 3.4 of][]{Tacchella2022}, except for the dust extinction $A_V$, where we used a uniform prior $0<A_V<8$ to match that used for the  \emph{BAGPIPES} and \emph{CIGALE} runs above.
\\ \newline

%-----------------------------------------------------------------
   \begin{figure*}
   \centering 
   \includegraphics[width=\textwidth]{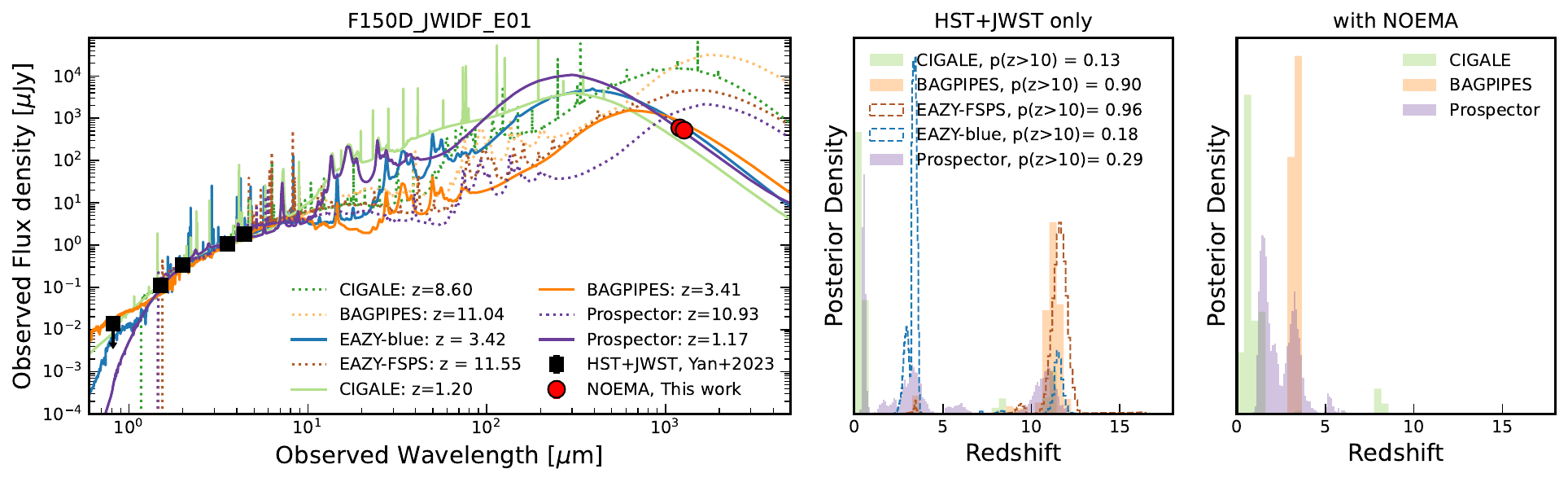}
   \includegraphics[width=\textwidth]{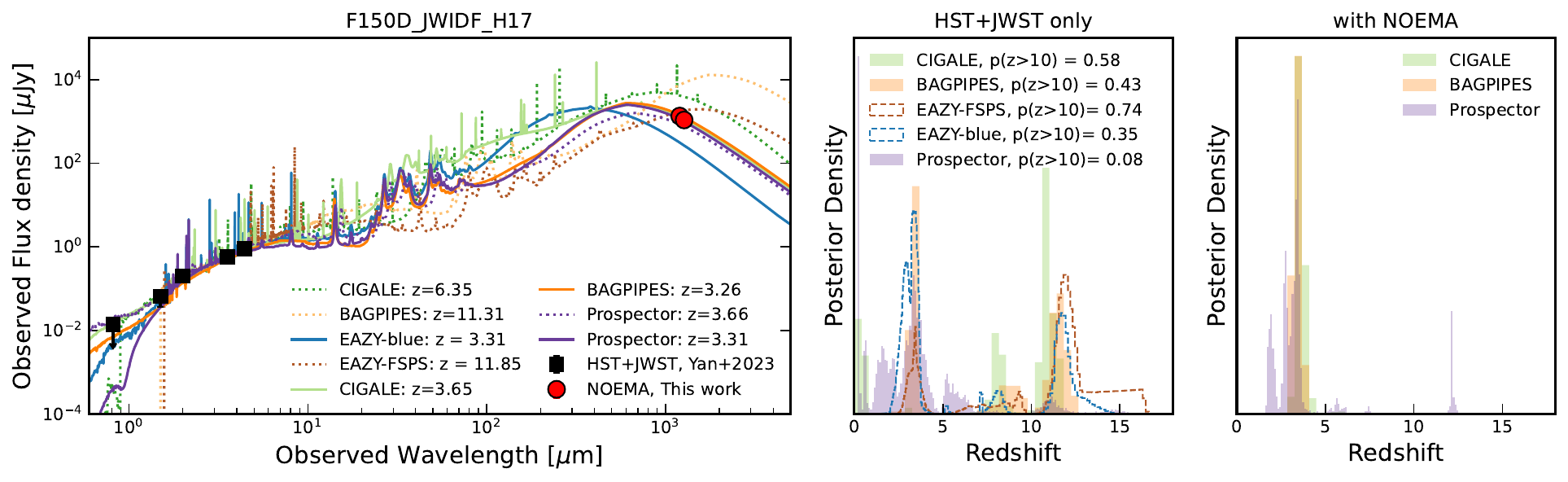}
   \includegraphics[width=\textwidth]{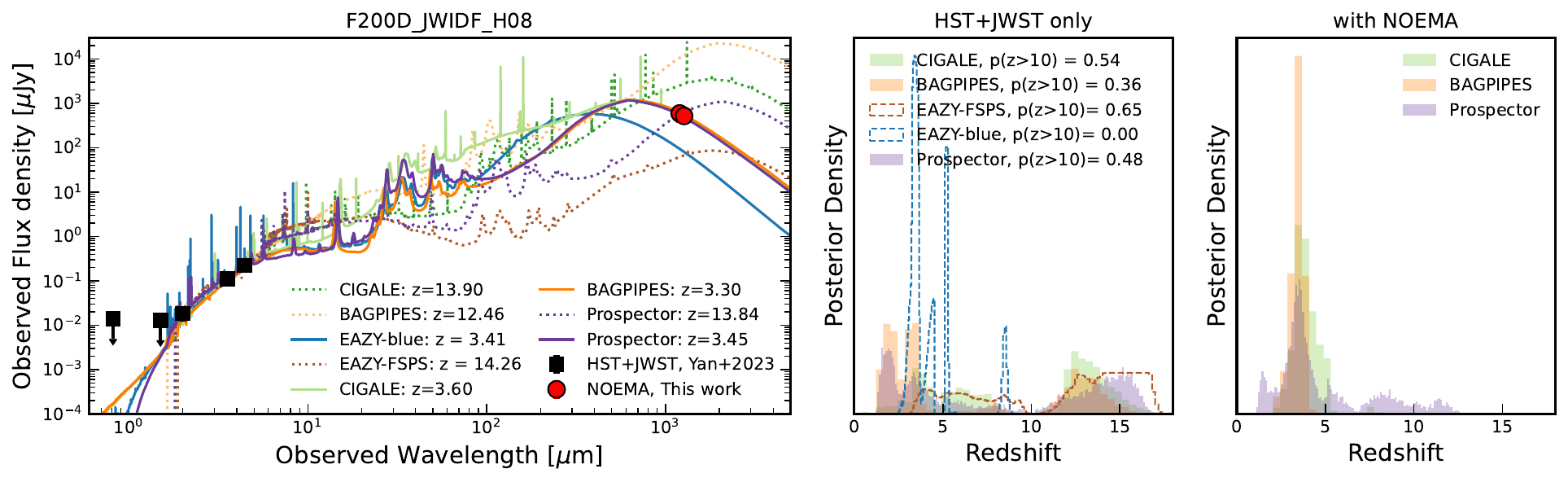}
   \includegraphics[width=\textwidth]{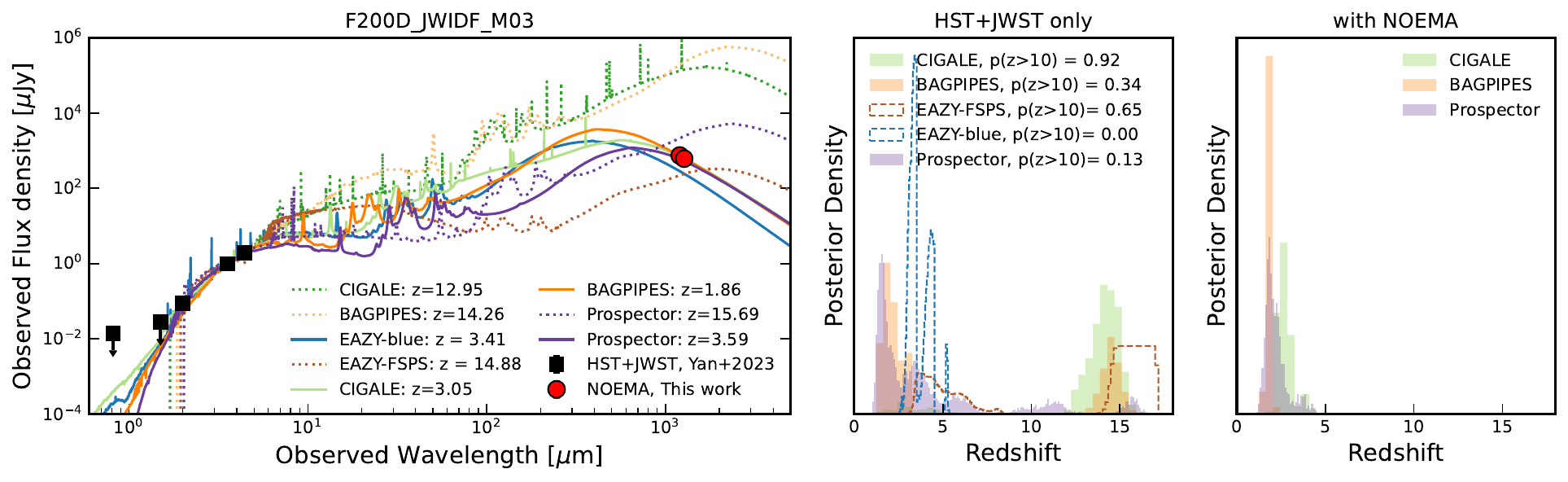}
      \caption{ SED and posterior redshift distributions of the four red 1.5-2.0 $\mu \rm{m}$ dropouts studied in this work. The best high-redshift SED using only the JWST+HST photometry are shown  with dotted lines and the best-fit low-redshift solutions found including the NOEMA data are shown with full lines (left). In the absence of FIR constraints, the four JWST bands and the one HST (black squares and limits) band cannot constrain the redshift distribution and allow or prefer a $z>10$ for most sources and codes (middle). The addition of the FIR constraints (red circles) break the degeneracies and clearly reveal the $z\sim 2-4$ nature of these sources  (right). Upper limits are shown at the two $\sigma$ level.}
         \label{fig:sed_pz}
   \end{figure*}
%
%______________________________________________________________

We fit the photometry of each object (see Table \ref{table:1}) with and without our NOEMA 1mm constraints. We show in Figure \ref{fig:sed_pz} the best-fit SEDs and the posterior redshift distributions. 
The first result is that a majority of the codes, in the absence of FIR constraints, either prefer a high-redshift ($z>6$) solution or allow one (with a $z>10$ solution peaking at least at half the likelihood of the intermediate redshift solution). The notable exception is \emph{EAZY} using the new \emph{blue\_shfz\_13} templates, which consistently prefers the lower redshift solution, although the posterior redshift distribution for F150W dropouts still contains solutions at $z>10$. The different performance of the \emph{blue\_shfz\_13} templates can be explained by the inferred rest-frame UV slopes of our targets if they would be at $z>10$. To derive the inferred rest-frame UV $\beta$ slopes, we fit the rest-frame UV ($1265-2580 \AA$) of all the  $z>10$ models from BAGPIPES, with a power law of $f_\lambda\propto \lambda^\beta$. We find UV slopes for the $z>10$ solutions $\beta = 0.13_{-0.13}^{+0.11}, -0.12_{-0.16}^{+0.17}, 1.28_{-0.36}^{+0.21},1.44_{-0.12}^{+0.12}$, for F150D\_JWIDF\_E01, F150D\_JWIDF\_H17, F200D\_JWIDF\_H08, and F200D\_JWIDF\_M03, respectively. These $\beta$ values are much higher than that the typical $-3\lesssim\beta\lesssim-1$ measured in confirmed $z>10$ galaxies \citep[e.g.][]{Bunker2023_GNz11, Curtis-Lake2023}. The absence of extremely red templates at high-redshift in the \emph{blue\_shfz\_13} set therefore explains the lower photometric redshift solutions for this particular \emph{EAZY} run.

\emph{Prospector} also prefers lower-redshift solutions, although the probability for a $z>10$ is still significant ($p(z>10) \sim 0.1-0.4$, see Fig. \ref{fig:sed_pz}). However, we note  that the \emph{Prospector} results heavily depend on the choice of priors for the various physical parameters. Indeed,  the maximum-likelihood solutions (plotted in dotted purple in Fig. \ref{fig:sed_pz}) are $z_{\rm{ML}}= 10.9, 3.66, 13.8, 15.6$ for F150D\_JWIDF\_E01, F150D\_JWIDF\_H17, F200D\_JWIDF\_H08, and F200D\_JWIDF\_M03, respectively. We also find the posterior redshift distribution to be strongly dependent on the dust parameters priors, for instance, a lognormal prior for the dust extinction will result in $z>10$ solution being strongly preferred. The \emph{Prospector} results are likely prior-dominated due to the low number of continuum datapoints.

The second important result is that the addition of the NOEMA observations completely transforms the posterior redshift distribution, as also shown in \citet{Zavala2023_z5dusty}. Indeed, once the FIR continuum constraints are included in the fits, an intermediate redshift solution ($z\sim2-4$) is strongly preferred in all codes.\footnote{We did not fit the full optical-to-FIR SED with \emph{EAZY} as the accuracy of the photo-z has not been tested with FIR data, even though the template extend to that regime (private communication, G. Brammer). In our tests we indeed find that the inclusion of constraints in the mm-regime do not change the photometric redshifts inferred.}.

\renewcommand{\arraystretch}{1.5} 
\begin{table}

    \caption{Key physical properties of the four 1.5-2.0 $\mu\rm{m}$ red dropouts.}
    \centering
    \small
    \begin{tabular}{ccccc}
    \hline \hline
    ID & $z$ & $\rm{SFR}$  & $\log M_*$   & $A_V$  \\ &  & $[M_\odot\ \rm{yr}^{-1}]$  & $[M_\odot]$   & [mag] \\ \hline 
          &\multicolumn{4}{c}{BAGPIPES} \\ \cline{2-5}
    F200D\_M03 & $1.80^{+0.12}_{-0.12}$ & $21^{+2}_{-3}$ &  $10.24^{+0.10}_{-0.20}$ &  $7.0^{+0.5}_{-0.4}$ \\
    F200D\_H08  & $3.46^{+0.24}_{-0.37}$ & $15^{+4}_{-3}$ &  $9.73^{+0.15}_{-0.17}$ & $3.8^{+0.8}_{-0.4}$ \\
    F150D\_H17  &  $3.37^{+0.21}_{-0.13}$ & $39^{+3}_{-3}$  & $9.97^{+0.10}_{-0.10}$ & $2.6^{+0.2}_{-0.2}$ \\
    F150D\_E01  & $3.29^{+0.12}_{-0.10}$ & $33^{+6}_{-6}$  & $10.49^{+0.06}_{-0.07}$& $2.2^{+0.1}_{-0.1}$  \\ \hline
    & \multicolumn{4}{c}{CIGALE} \\ \cline{2-5}
 F200D\_M03 &  $2.75\pm0.43$ & $47\pm21$ & $10.57\pm0.37$ & $3.9\pm0.7$ \\
 F200D\_H08 &  $3.89\pm0.75$ & $42\pm16$  & $9.86\pm0.38$ & $3.1\pm0.6$ \\
 F150D\_H17 & $3.63\pm0.28$ & $117\pm18$ & $9.94\pm0.17$ & $2.0\pm0.1$ \\
F150D\_E01 &$ 1.11^{+1.53}_{-1.10}$ & $15^{+19}_{-15}$ & $8.79\pm0.64$ &  $5.0\pm1.7$  \\
    \end{tabular}
    \tablefoot{The properties are derived using the full optical to FIR SED constraints. For each parameter we give the median value and error from the 16th, 50th, and 84th percentiles of the BAGPIPES posterior and CIGALE's Bayesian estimates.}
    \label{table:2}
\end{table}

\begin{figure}
    \centering
    \includegraphics[width=0.49\textwidth]{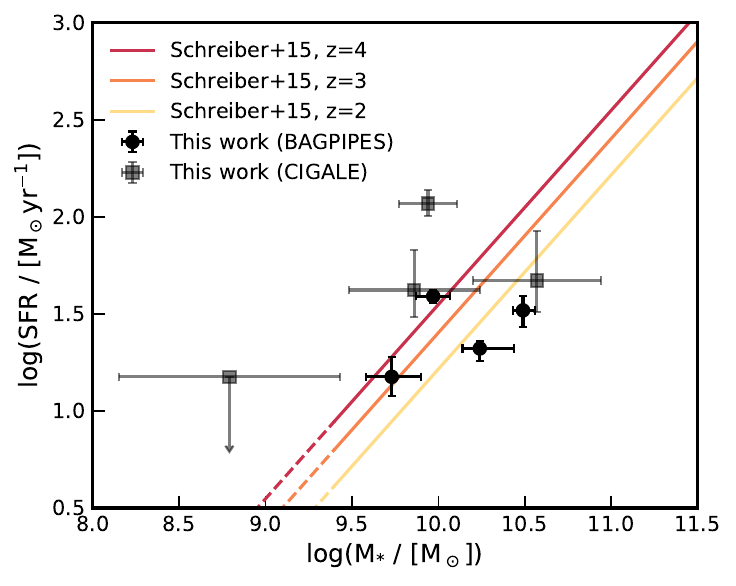} 
    \caption{Inferred star formation rate against stellar mass for the galaxies studied in this work. Results from \emph{BAGPIPES} are shown in black circles, whereas those from \emph{CIGALE} are shown in grey squares. The SFR upper limit (F200D\_JWIDF\_M03 fitted with CIGALE) is shown at the $2\sigma$ level. The coloured lines show the galaxy main sequence at $z=2,3,4$ from \citet{Schreiber2015}.}
    \label{fig:ms}
\end{figure}

Using \emph{BAGPIPES} and \emph{CIGALE} to fit their optical-to-FIR SED, we find that the four sources are galaxies at intermediate redshift ($z\sim2-4$), with star formation rates of SFR$\sim 20-150 M_\odot\ \rm{yr}^{-1}$, stellar masses of $ 9.7 < \log M_* / [M_\odot] < 11.4$, and high obscuration ($ 2\lesssim A_V\lesssim 7$). \emph{CIGALE} tends to prefer solutions with lower redshifts, along with higher masses and dust attenuations, but the \emph{BAGPIPES} and \emph{CIGALE} results are consistent within the $ 1\sigma $ errors (see further Table \ref{table:2} for the full results of the two codes). The most important conclusion is that these objects are obscured galaxies at $z\sim2-4$ that lie on the galaxy main sequence (see Fig. \ref{fig:ms}), similarly to the HST-faint/JWST-detected objects reported by \citet{Barrufet2023}.

\begin{figure}
    \centering
    \includegraphics[width=0.49\textwidth]{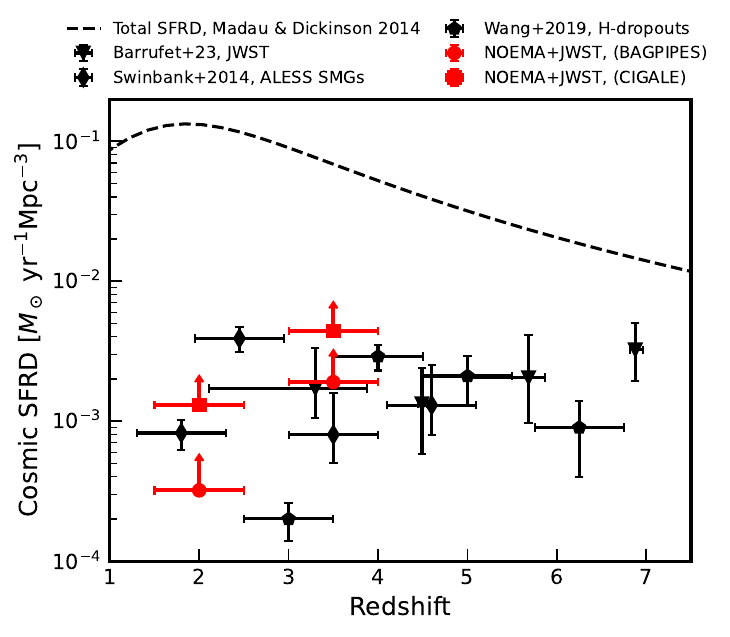}
    \caption{Cosmic star formation rate density contribution from dusty star-forming galaxies. The constraints from this work sample are shown with red squares (BAGPIPES) and circles (CIGALE). Selected constraints from the litterature are shown in black \citep[][but see further text]{Swinbank2014,TWang2019,Barrufet2023}. We also show the total SFRD with a dashed black line \citep{Madau2014}. The extremely red JWST dropouts represent $\gtrsim 5\%$ of the CSFRD at $z\sim 3-4$}
    \label{fig:csfrd}
\end{figure}

We used the best-fit \emph{BAGPIPES}  spectrum to derive the obscured and unobscured star formation rate, using the \citet{Kennicutt2012} conversion factors from the rest-frame FIR ($3-1100\ \mu\rm{m}$) and UV ($1550\ \AA$) luminosities. We find that that all galaxies are highly obscured, with a negligible contribution of the unobscured SFR to the total SFR, with $\rm{SFR_{UV}} / (\rm{SFR}_{UV}+\rm{SFR}_{IR}) \sim 10^{-7} -  10^{-3}$. We also estimate the contribution of these objects to the total cosmic star formation rate density using the median SFR of the BAGPIPES/CIGALE posterior and the area of the PEARLS-IDF field our targets were selected from ($14.2$ arcmin$^{2}$). Using a wide redshift bin $\Delta z=1$ and ignoring the effect of the unknown selection function, we find lower limits on the cosmic SFR density (CSFRD) of $\phi= (3.2-13.0)\times10^{-4}\ M_\odot\ \rm{yr}^{-1} \ \rm{cMpc}^{-3}$ at $z=2$ and $\phi=(1.9-4.4)\times10^{-3}\ M_\odot\ \rm{yr}^{-1} \ \rm{cMpc}^{-3}$ at $z=3.5$, where the range of values encompasses the scatter between the CIGALE and BAGPIPES results (see Table \ref{table:2}). The CSFRD we derive for the four sources is in good agreement with the literature constraints on the contribution of obscured galaxies to the CSFRD at these redshifts \citep[see Fig. \ref{fig:csfrd}, right panel; and e.g.][]{Blain1999,Chapman2005,Barger2012,Swinbank2014,TWang2019, Dudzeviciute2020,Shu2022, Barrufet2023}, representing a $>0.2-1.0\ (>3-6\%)$ contribution to the total CSFRD at $z=2(3.5)$.

\section{Conclusions}

We present a NOEMA follow-up study of red 1.5--2.0\,$\mu\rm{m}$ dropouts in JWST PEARLS-IDF, showing that sub-mm follow-up is highly efficient at unveiling their true nature. Their FIR continuum emission is detected at a signal-to-noise ration of S/N$\sim 15$ in our the 236, 252 GHz observations, with typical values of $0.5-1.5\ \rm{mJy}$. Our SED modelling shows that such a faint continuum in the mm-regime unambiguously argues against an ultra-high-redshift ($z>10$) nature (Fig. \ref{fig:predictions_redshift} and Appendix \ref{appendix:A}). Instead, the 1.5--2.0\,$\mu\rm{m}$ breaks and red JWST colours are shown to select massive, star-forming, and dust-obscured galaxies at $z\sim 2-4$, extending the main sequence of SMGs and optically faint galaxies. These objects contribute a non-negligible fraction of the CSFRD at $z\simeq 3.5$ ($\phi\simeq (1.9-4.4)\times10^{-3}\ M_\odot\ \rm{yr}^{-1} \ \rm{cMpc}^{-3}$), as found in previous studies specifically targeting optically faint `HST-dark' or DSFGs galaxies. This demonstrates the efficiency of combining mm continuum observations with infrared imaging in a limited number of JWST filters to probe obscured star-formation beyond cosmic noon.

This work show specific cases where a simple $\sim 1-2\ \mu \rm{m}$ break selection is prone to select optically faint galaxies instead of ultra-high-redshift galaxies when no further criteria on the UV slopes of star-formation rate histories have been applied. Fortunately, we have demonstrated that red objects can be robustly identified with observations of their rest-frame FIR \citep[see also][]{Zavala2023_z5dusty}. The expected peak FIR continuum of $\sim 1\ \rm{mJy}$ (and the characteristic continuum slope; see Fig. \ref{fig:predictions_redshift}) can be detected at $10\sigma$ in $\sim 3(20)\ $ min of on-source time with ALMA (NOEMA), representing a highly efficient alternative to spectroscopic follow-up with JWST. We note that this is a different situation to previous works focusing on JWST high-redshift objects with blue colours. There, expensive redshift looking for rest-frame FIR emission lines must be conducted as the continuum is likely too faint to be detected and does not significantly constrain the redshift of the galaxy \citep[e.g.][]{Kaasinen2023,Bakx2023, Popping2023,Fujimoto2023}.

We have also demonstrated that the use of bluer templates at higher redshift, such as those made recently available with \emph{EAZY} (\emph{blue\_sfhz\_13}), improve the photo-z selection and characterisation of red dropouts in the absence of observations in the mm regime. However, we note  that these templates find a non-negligible high-redshift probability for the two F150W dropouts (e.g. $p(z>10) = 0.18$  and $0.35$, see Fig. \ref{fig:sed_pz}). Similarly, \emph{Prospector} is found to prefer lower-redshift galaxies even in the absence of FIR constraints. Yet it also gives broad posterior redshift distributions, which allow for a $z>10$ solution, even when including our NOEMA constraints. Whilst some studies have already rejected such objects with a substantial low-redshift probability from $z\gtrsim 10$ samples with a variety of additional criteria \citep[e.g.][]{Finkelstein2023, Finkelstein2023a, Harikane2023}, further tests of these templates and redshift priors with statistical samples of spectroscopically-confirmed red JWST-only galaxies are still necessary to improve the robustness of photometric redshifts. Another unexplored avenue is the use of morphology priors, as in the case of the red objects studied in this work or those in \citet[][]{Nelson2022}, for instance, which have elongated or disky morphologies; these are evidently in stark contrast with the almost point-like $z>10$ sources confirmed so far. Finally, for surveys with a large enough field of view, a CSFRD prior could be imposed to guide the ensemble prediction for all galaxies in the field (see more in the appendix \label{appendix:A}). 

In summary, distinguishing obscured star-forming galaxies from ultra-high-redshift galaxies with sparse JWST wavelength coverage is difficult when using standard SED codes (templates), although dedicated templates and/or appropriate priors could improve some of the photometric redshifts. However, follow-up (sub-)mm observations of the FIR continuum provide a highly efficient way to determine the nature of high-redshift dusty star-forming galaxies.

\begin{acknowledgements}
The authors thank the anonymous referee for comments and suggestions which improved this Letter. RAM thanks the IRAM support staff and in particular J. Orkisz for the observations and data reduction.
RAM, LAB, FW acknowledge support from the ERC Advanced Grant 740246 (Cosmic\_Gas). 
RAM, PO acknowledge support from the Swiss National Science Foundation (SNSF) through project grant 200020\_207349.
\\
This work is based on observations carried out under project number W22EG with the IRAM NOEMA Interferometer. IRAM is supported by INSU/CNRS (France), MPG (Germany) and IGN (Spain). \\

This work made use of the following Python packages: \emph{numpy} \citep{Numpy2020}, \emph{matplotlib} \citep{Hunter2007}, \emph{scipy} \citep{Virtanen2020}, \emph{Astropy} \citep{astropy:2013, astropy:2018, astropy:2022}, \emph{interferopy} \citep*{interferopy}, \emph{BAGPIPES} \citep{Carnall2018} , \emph{EAZY} \citet{Brammer2012}, \emph{CIGALE} \citet{Boquien2019}, and \emph{PROSPECTOR} \citet{Johnson2021}.

\end{acknowledgements}


\bibliographystyle{./bibtex/aa}
\bibliography{library_AAL}

\begin{thebibliography}{68}
\expandafter\ifx\csname natexlab\endcsname\relax\def\natexlab#1{#1}\fi

\bibitem[{Adams {et~al.}(2022)Adams, Conselice, Ferreira, Austin, Trussler,
  Juodžbalis, Wilkins, Caruana, Dayal, Verma, \& Vijayan}]{Adams2022a}
Adams, N.~J., Conselice, C.~J., Ferreira, L., {et~al.} 2022, Monthly Notices of
  the Royal Astronomical Society, 518, 4755

\bibitem[{{Alcalde Pampliega} {et~al.}(2019){Alcalde Pampliega},
  Pérez-González, Barro, Sánchez, Eliche-Moral, Cardiel, Hernán-Caballero,
  Rodriguez-Muñoz, Blázquez, \& Esquej}]{Alcalde-Pampliega2019}
{Alcalde Pampliega}, B., Pérez-González, P.~G., Barro, G., {et~al.} 2019, The
  Astrophysical Journal, 876, 135

\bibitem[{{Arrabal Haro} {et~al.}(2023{\natexlab{a}}){Arrabal Haro}, Dickinson,
  Finkelstein, Fujimoto, Fernández, Kartaltepe, Jung, Cole, Burgarella,
  Chworowsky, Hutchison, Morales, Papovich, Simons, Amorín, Backhaus, Bagley,
  Bisigello, Calabrò, Castellano, Cleri, Davé, Dekel, Ferguson, Fontana,
  Gawiser, Giavalisco, Harish, Hathi, Hirschmann, Holwerda, Huertas-Company,
  Koekemoer, Larson, Lucas, Mobasher, Pérez-González, Pirzkal, Rose, Santini,
  Trump, de~la Vega, Wang, Weiner, Wilkins, Yang, Yung, Zavala, \&
  Woods}]{ArrabalHaro2023}
{Arrabal Haro}, P., Dickinson, M., Finkelstein, S.~L., {et~al.}
  2023{\natexlab{a}}, The Astrophysical Journal Letters, 951, L22

\bibitem[{{Arrabal Haro} {et~al.}(2023{\natexlab{b}}){Arrabal Haro},
  {Dickinson}, {Finkelstein}, {Kartaltepe}, {Donnan}, {Burgarella}, {Carnall},
  {Cullen}, {Dunlop}, {Fern{\'a}ndez}, {Fujimoto}, {Jung}, {Krips}, {Larson},
  {Papovich}, {P{\'e}rez-Gonz{\'a}lez}, {Amor{\'\i}n}, {Bagley}, {Buat},
  {Casey}, {Chworowsky}, {Cohen}, {Ferguson}, {Giavalisco}, {Huertas-Company},
  {Hutchison}, {Kocevski}, {Koekemoer}, {Lucas}, {McLeod}, {McLure}, {Pirzkal},
  {Seill{\'e}}, {Trump}, {Weiner}, {Wilkins}, \& {Zavala}}]{ArrabalHaro2023a}
{Arrabal Haro}, P., {Dickinson}, M., {Finkelstein}, S.~L., {et~al.}
  2023{\natexlab{b}}, \nat, 622, 707

\bibitem[{{Astropy Collaboration} {et~al.}(2022){Astropy Collaboration},
  {Price-Whelan}, {Lim}, {Earl}, {Starkman}, {Bradley}, {Shupe}, {Patil},
  {Corrales}, {Brasseur}, {N{"o}the}, {Donath}, {Tollerud}, {Morris},
  {Ginsburg}, {Vaher}, {Weaver}, {Tocknell}, {Jamieson}, {van Kerkwijk},
  {Robitaille}, {Merry}, {Bachetti}, {G{"u}nther}, {Aldcroft},
  {Alvarado-Montes}, {Archibald}, {B{'o}di}, {Bapat}, {Barentsen}, {Baz{'a}n},
  {Biswas}, {Boquien}, {Burke}, {Cara}, {Cara}, {Conroy}, {Conseil}, {Craig},
  {Cross}, {Cruz}, {D'Eugenio}, {Dencheva}, {Devillepoix}, {Dietrich},
  {Eigenbrot}, {Erben}, {Ferreira}, {Foreman-Mackey}, {Fox}, {Freij}, {Garg},
  {Geda}, {Glattly}, {Gondhalekar}, {Gordon}, {Grant}, {Greenfield}, {Groener},
  {Guest}, {Gurovich}, {Handberg}, {Hart}, {Hatfield-Dodds}, {Homeier},
  {Hosseinzadeh}, {Jenness}, {Jones}, {Joseph}, {Kalmbach}, {Karamehmetoglu},
  {Ka{l}uszy{'n}ski}, {Kelley}, {Kern}, {Kerzendorf}, {Koch}, {Kulumani},
  {Lee}, {Ly}, {Ma}, {MacBride}, {Maljaars}, {Muna}, {Murphy}, {Norman},
  {O'Steen}, {Oman}, {Pacifici}, {Pascual}, {Pascual-Granado}, {Patil},
  {Perren}, {Pickering}, {Rastogi}, {Roulston}, {Ryan}, {Rykoff}, {Sabater},
  {Sakurikar}, {Salgado}, {Sanghi}, {Saunders}, {Savchenko}, {Schwardt},
  {Seifert-Eckert}, {Shih}, {Jain}, {Shukla}, {Sick}, {Simpson},
  {Singanamalla}, {Singer}, {Singhal}, {Sinha}, {Sip{H{o}}cz}, {Spitler},
  {Stansby}, {Streicher}, {{{S}}umak}, {Swinbank}, {Taranu}, {Tewary},
  {Tremblay}, {Val-Borro}, {Van Kooten}, {Vasovi{'c}}, {Verma}, {de Miranda
  Cardoso}, {Williams}, {Wilson}, {Winkel}, {Wood-Vasey}, {Xue}, {Yoachim},
  {Zhang}, {Zonca}, \& {Astropy Project Contributors}}]{astropy:2022}
{Astropy Collaboration}, {Price-Whelan}, A.~M., {Lim}, P.~L., {et~al.} 2022,
  \apj, 935, 167

\bibitem[{{Astropy Collaboration} {et~al.}(2018){Astropy Collaboration},
  {Price-Whelan}, {Sip{\H{o}}cz}, {G{\"u}nther}, {Lim}, {Crawford}, {Conseil},
  {Shupe}, {Craig}, {Dencheva}, {Ginsburg}, {Vand erPlas}, {Bradley},
  {P{\'e}rez-Su{\'a}rez}, {de Val-Borro}, {Aldcroft}, {Cruz}, {Robitaille},
  {Tollerud}, {Ardelean}, {Babej}, {Bach}, {Bachetti}, {Bakanov}, {Bamford},
  {Barentsen}, {Barmby}, {Baumbach}, {Berry}, {Biscani}, {Boquien}, {Bostroem},
  {Bouma}, {Brammer}, {Bray}, {Breytenbach}, {Buddelmeijer}, {Burke},
  {Calderone}, {Cano Rodr{\'\i}guez}, {Cara}, {Cardoso}, {Cheedella}, {Copin},
  {Corrales}, {Crichton}, {D'Avella}, {Deil}, {Depagne}, {Dietrich}, {Donath},
  {Droettboom}, {Earl}, {Erben}, {Fabbro}, {Ferreira}, {Finethy}, {Fox},
  {Garrison}, {Gibbons}, {Goldstein}, {Gommers}, {Greco}, {Greenfield},
  {Groener}, {Grollier}, {Hagen}, {Hirst}, {Homeier}, {Horton}, {Hosseinzadeh},
  {Hu}, {Hunkeler}, {Ivezi{\'c}}, {Jain}, {Jenness}, {Kanarek}, {Kendrew},
  {Kern}, {Kerzendorf}, {Khvalko}, {King}, {Kirkby}, {Kulkarni}, {Kumar},
  {Lee}, {Lenz}, {Littlefair}, {Ma}, {Macleod}, {Mastropietro}, {McCully},
  {Montagnac}, {Morris}, {Mueller}, {Mumford}, {Muna}, {Murphy}, {Nelson},
  {Nguyen}, {Ninan}, {N{\"o}the}, {Ogaz}, {Oh}, {Parejko}, {Parley}, {Pascual},
  {Patil}, {Patil}, {Plunkett}, {Prochaska}, {Rastogi}, {Reddy Janga},
  {Sabater}, {Sakurikar}, {Seifert}, {Sherbert}, {Sherwood-Taylor}, {Shih},
  {Sick}, {Silbiger}, {Singanamalla}, {Singer}, {Sladen}, {Sooley},
  {Sornarajah}, {Streicher}, {Teuben}, {Thomas}, {Tremblay}, {Turner},
  {Terr{\'o}n}, {van Kerkwijk}, {de la Vega}, {Watkins}, {Weaver}, {Whitmore},
  {Woillez}, {Zabalza}, \& {Astropy Contributors}}]{astropy:2018}
{Astropy Collaboration}, {Price-Whelan}, A.~M., {Sip{\H{o}}cz}, B.~M., {et~al.}
  2018, \aj, 156, 123

\bibitem[{{Astropy Collaboration} {et~al.}(2013){Astropy Collaboration},
  {Robitaille}, {Tollerud}, {Greenfield}, {Droettboom}, {Bray}, {Aldcroft},
  {Davis}, {Ginsburg}, {Price-Whelan}, {Kerzendorf}, {Conley}, {Crighton},
  {Barbary}, {Muna}, {Ferguson}, {Grollier}, {Parikh}, {Nair}, {Unther},
  {Deil}, {Woillez}, {Conseil}, {Kramer}, {Turner}, {Singer}, {Fox}, {Weaver},
  {Zabalza}, {Edwards}, {Azalee Bostroem}, {Burke}, {Casey}, {Crawford},
  {Dencheva}, {Ely}, {Jenness}, {Labrie}, {Lim}, {Pierfederici}, {Pontzen},
  {Ptak}, {Refsdal}, {Servillat}, \& {Streicher}}]{astropy:2013}
{Astropy Collaboration}, {Robitaille}, T.~P., {Tollerud}, E.~J., {et~al.} 2013,
  \aap, 558, A33

\bibitem[{{Bakx} {et~al.}(2023){Bakx}, {Zavala}, {Mitsuhashi}, {Treu},
  {Fontana}, {Tadaki}, {Casey}, {Castellano}, {Glazebrook}, {Hagimoto},
  {Ikeda}, {Jones}, {Leethochawalit}, {Mason}, {Morishita}, {Nanayakkara},
  {Pentericci}, {Roberts-Borsani}, {Santini}, {Serjeant}, {Tamura}, {Trenti},
  \& {Vanzella}}]{Bakx2023}
{Bakx}, T. J.~L.~C., {Zavala}, J.~A., {Mitsuhashi}, I., {et~al.} 2023, \mnras,
  519, 5076

\bibitem[{{Barger} \& {Cowie}(2023)}]{Barger2023}
{Barger}, A.~J. \& {Cowie}, L.~L. 2023, \apj, 956, 95

\bibitem[{{Barger} {et~al.}(2012){Barger}, {Wang}, {Cowie}, {Owen}, {Chen}, \&
  {Williams}}]{Barger2012}
{Barger}, A.~J., {Wang}, W.~H., {Cowie}, L.~L., {et~al.} 2012, \apj, 761, 89

\bibitem[{Barrufet {et~al.}(2023)Barrufet, Oesch, Weibel, Brammer, Bezanson,
  Bouwens, Fudamoto, Gonzalez, Gottumukkala, Illingworth, Heintz, Holden,
  Labbe, Magee, Naidu, Nelson, Stefanon, Smit, van Dokkum, Weaver, \&
  Williams}]{Barrufet2023}
Barrufet, L., Oesch, P.~A., Weibel, A., {et~al.} 2023, Monthly Notices of the
  Royal Astronomical Society, 522, 449

\bibitem[{{Blain} {et~al.}(1999){Blain}, {Smail}, {Ivison}, \&
  {Kneib}}]{Blain1999}
{Blain}, A.~W., {Smail}, I., {Ivison}, R.~J., \& {Kneib}, J.~P. 1999, \mnras,
  302, 632

\bibitem[{Boogaard {et~al.}(2021)Boogaard, Meyer, \& Novak}]{interferopy}
Boogaard, L., Meyer, R.~A., \& Novak, M. 2021, {Interferopy: analysing
  datacubes from radio-to-submm observations}, 10.5281/ZENODO.5775603

\bibitem[{Boquien {et~al.}(2019)Boquien, Burgarella, Roehlly, Buat, Ciesla,
  Corre, Inoue, \& Salas}]{Boquien2019}
Boquien, M., Burgarella, D., Roehlly, Y., {et~al.} 2019, Astronomy \&
  Astrophysics, 622, A103

\bibitem[{Bouwens {et~al.}(2023)Bouwens, Stefanon, Brammer, Oesch,
  Herard-Demanche, Illingworth, Matthee, Naidu, Dokkum, \&
  Leeuwen}]{Bouwens2023a}
Bouwens, R.~J., Stefanon, M., Brammer, G., {et~al.} 2023, Monthly Notices of
  the Royal Astronomical Society, 523, 1036

\bibitem[{Brammer {et~al.}(2012)Brammer, van Dokkum, Franx, Fumagalli, Patel,
  Rix, Skelton, Kriek, Nelson, Schmidt, Bezanson, da~Cunha, Erb, Fan,
  Schreiber, Illingworth, Labbé, Leja, Lundgren, Magee, Marchesini, McCarthy,
  Momcheva, Muzzin, Quadri, Steidel, Tal, Wake, Whitaker, \&
  Williams}]{Brammer2012}
Brammer, G.~B., van Dokkum, P.~G., Franx, M., {et~al.} 2012, The Astrophysical
  Journal Supplement Series, 200, 13

\bibitem[{Bruzual \& Charlot(2003)}]{Bruzual2003}
Bruzual, G. \& Charlot, S. 2003, Monthly Notices of the Royal Astronomical
  Society, 344, 1000

\bibitem[{Bunker {et~al.}(2023)Bunker, Cameron, Curtis-Lake, Jakobsen,
  Carniani, Curti, Witstok, Maiolino, D'Eugenio, Looser, Willott, Bonaventura,
  Hainline, Uebler, Willmer, Saxena, Smit, Alberts, Arribas, Baker, Baum,
  Bhatawdekar, Bowler, Boyett, Charlot, Chen, Chevallard, Circosta, DeCoursey,
  de~Graaff, Egami, Eisenstein, Endsley, Ferruit, Giardino, Hausen, Helton,
  Hviding, Ji, Johnson, Jones, Kumari, Laseter, Luetzgendorf, Maseda, Nelson,
  Parlanti, Perna, Rawle, Rix, Rieke, Robertson, Pino, Sandles, Scholtz,
  Sharpe, Skarbinski, Stark, Sun, Tacchella, Topping, Villanueva, Wallace,
  Williams, \& Woodrum}]{Bunker2023_JADESrelease}
Bunker, A.~J., Cameron, A.~J., Curtis-Lake, E., {et~al.} 2023

\bibitem[{{Bunker} {et~al.}(2023){Bunker}, {Saxena}, {Cameron}, {Willott},
  {Curtis-Lake}, {Jakobsen}, {Carniani}, {Smit}, {Maiolino}, {Witstok},
  {Curti}, {D'Eugenio}, {Jones}, {Ferruit}, {Arribas}, {Charlot}, {Chevallard},
  {Giardino}, {de Graaff}, {Looser}, {L{\"u}tzgendorf}, {Maseda}, {Rawle},
  {Rix}, {Del Pino}, {Alberts}, {Egami}, {Eisenstein}, {Endsley}, {Hainline},
  {Hausen}, {Johnson}, {Rieke}, {Rieke}, {Robertson}, {Shivaei}, {Stark},
  {Sun}, {Tacchella}, {Tang}, {Williams}, {Willmer}, {Baker}, {Baum},
  {Bhatawdekar}, {Bowler}, {Boyett}, {Chen}, {Circosta}, {Helton}, {Ji},
  {Kumari}, {Lyu}, {Nelson}, {Parlanti}, {Perna}, {Sandles}, {Scholtz},
  {Suess}, {Topping}, {{\"U}bler}, {Wallace}, \& {Whitler}}]{Bunker2023_GNz11}
{Bunker}, A.~J., {Saxena}, A., {Cameron}, A.~J., {et~al.} 2023, \aap, 677, A88

\bibitem[{Calzetti {et~al.}(2000)Calzetti, Armus, Bohlin, Kinney, Koornneef, \&
  Storchi‐Bergmann}]{Calzetti2000}
Calzetti, D., Armus, L., Bohlin, R.~C., {et~al.} 2000, The Astrophysical
  Journal, 533, 682

\bibitem[{Carnall {et~al.}(2018)Carnall, McLure, Dunlop, \&
  Davé}]{Carnall2018}
Carnall, A.~C., McLure, R.~J., Dunlop, J.~S., \& Davé, R. 2018, Monthly
  Notices of the Royal Astronomical Society, 480, 4379

\bibitem[{Casey {et~al.}(2018)Casey, Hodge, Zavala, Spilker, da~Cunha, Staguhn,
  Finkelstein, \& Drew}]{Casey2018}
Casey, C.~M., Hodge, J., Zavala, J.~A., {et~al.} 2018, The Astrophysical
  Journal, 862, 78

\bibitem[{Casey {et~al.}(2014)Casey, Narayanan, \& Cooray}]{Casey2014}
Casey, C.~M., Narayanan, D., \& Cooray, A. 2014, Physics Reports, 541, 45

\bibitem[{Chabrier(2003)}]{Chabrier2003}
Chabrier, G. 2003, Publications of the Astronomical Society of the Pacific,
  115, 763

\bibitem[{{Chapman} {et~al.}(2005){Chapman}, {Blain}, {Smail}, \&
  {Ivison}}]{Chapman2005}
{Chapman}, S.~C., {Blain}, A.~W., {Smail}, I., \& {Ivison}, R.~J. 2005, \apj,
  622, 772

\bibitem[{Curtis-Lake {et~al.}(2023)Curtis-Lake, Carniani, Cameron, Charlot,
  Jakobsen, Maiolino, Bunker, Witstok, Smit, Chevallard, Willott, Ferruit,
  Arribas, Bonaventura, Curti, D’Eugenio, Franx, Giardino, Looser,
  Lützgendorf, Maseda, Rawle, Rix, del Pino, Übler, Sirianni, Dressler,
  Egami, Eisenstein, Endsley, Hainline, Hausen, Johnson, Rieke, Robertson,
  Shivaei, Stark, Tacchella, Williams, Willmer, Bhatawdekar, Bowler, Boyett,
  Chen, de~Graaff, Helton, Hviding, Jones, Kumari, Lyu, Nelson, Perna, Sandles,
  Saxena, Suess, Sun, Topping, Wallace, \& Whitler}]{Curtis-Lake2023}
Curtis-Lake, E., Carniani, S., Cameron, A., {et~al.} 2023, Nature Astronomy
  2023 7:5, 7, 622

\bibitem[{Donnan {et~al.}(2022)Donnan, McLeod, Dunlop, McLure, Carnall, Begley,
  Cullen, Hamadouche, Bowler, Magee, McCracken, Milvang-Jensen, Moneti, \&
  Targett}]{Donnan2022}
Donnan, C.~T., McLeod, D.~J., Dunlop, J.~S., {et~al.} 2022, Monthly Notices of
  the Royal Astronomical Society, 518, 6011

\bibitem[{{Dudzevi{\v{c}}i{\={u}}t{\.{e}}}
  {et~al.}(2020){Dudzevi{\v{c}}i{\={u}}t{\.{e}}}, {Smail}, {Swinbank}, {Stach},
  {Almaini}, {da Cunha}, {An}, {Arumugam}, {Birkin}, {Blain}, {Chapman},
  {Chen}, {Conselice}, {Coppin}, {Dunlop}, {Farrah}, {Geach}, {Gullberg},
  {Hartley}, {Hodge}, {Ivison}, {Maltby}, {Scott}, {Simpson}, {Simpson},
  {Thomson}, {Walter}, {Wardlow}, {Weiss}, \& {van der
  Werf}}]{Dudzeviciute2020}
{Dudzevi{\v{c}}i{\={u}}t{\.{e}}}, U., {Smail}, I., {Swinbank}, A.~M., {et~al.}
  2020, \mnras, 494, 3828

\bibitem[{Elbaz {et~al.}(2011)Elbaz, Dickinson, Hwang, Díaz-Santos, Magdis,
  Magnelli, Borgne, Galliano, Pannella, Chanial, Armus, Charmandaris, Daddi,
  Aussel, Popesso, Kartaltepe, Altieri, Valtchanov, Coia, Dannerbauer, Dasyra,
  Leiton, Mazzarella, Alexander, Buat, Burgarella, Chary, Gilli, Ivison,
  Juneau, Floc'h, Lutz, Morrison, Mullaney, Murphy, Pope, Scott, Brodwin,
  Calzetti, Cesarsky, Charlot, Dole, Eisenhardt, Ferguson, Schreiber, Frayer,
  Giavalisco, Huynh, Koekemoer, Papovich, Reddy, Surace, Teplitz, \&
  Wilson}]{Elbaz2011}
Elbaz, D., Dickinson, M., Hwang, H.~S., {et~al.} 2011, A\&A, 533, 119

\bibitem[{{Finkelstein} {et~al.}(2023{\natexlab{a}}){Finkelstein}, {Bagley},
  {Ferguson}, {Wilkins}, {Kartaltepe}, {Papovich}, {Yung}, {Arrabal Haro},
  {Behroozi}, {Dickinson}, {Kocevski}, {Koekemoer}, {Larson}, {Le Bail},
  {Morales}, {P{\'e}rez-Gonz{\'a}lez}, {Burgarella}, {Dav{\'e}}, {Hirschmann},
  {Somerville}, {Wuyts}, {Bromm}, {Casey}, {Fontana}, {Fujimoto}, {Gardner},
  {Giavalisco}, {Grazian}, {Grogin}, {Hathi}, {Hutchison}, {Jha}, {Jogee},
  {Kewley}, {Kirkpatrick}, {Long}, {Lotz}, {Pentericci}, {Pierel}, {Pirzkal},
  {Ravindranath}, {Ryan}, {Trump}, {Yang}, {Bhatawdekar}, {Bisigello}, {Buat},
  {Calabr{\`o}}, {Castellano}, {Cleri}, {Cooper}, {Croton}, {Daddi}, {Dekel},
  {Elbaz}, {Franco}, {Gawiser}, {Holwerda}, {Huertas-Company}, {Jaskot},
  {Leung}, {Lucas}, {Mobasher}, {Pandya}, {Tacchella}, {Weiner}, \&
  {Zavala}}]{Finkelstein2023}
{Finkelstein}, S.~L., {Bagley}, M.~B., {Ferguson}, H.~C., {et~al.}
  2023{\natexlab{a}}, \apjl, 946, L13

\bibitem[{Finkelstein {et~al.}(2022)Finkelstein, Bagley, Haro, Dickinson,
  Ferguson, Kartaltepe, Papovich, Burgarella, Kocevski, Huertas-Company, Iyer,
  Koekemoer, Larson, Pérez-González, Rose, Tacchella, Wilkins, Chworowsky,
  Medrano, Morales, Somerville, Yung, Fontana, Giavalisco, Grazian, Grogin,
  Kewley, Kirkpatrick, Kurczynski, Lotz, Pentericci, Pirzkal, Ravindranath,
  Ryan, Trump, Yang, Almaini, Amorín, Annunziatella, Backhaus, Barro,
  Behroozi, Bell, Bhatawdekar, Bisigello, Bromm, Buat, Buitrago, Calabrò,
  Casey, Castellano, Óscar A.~Chávez~Ortiz, Ciesla, Cleri, Cohen, Cole,
  Cooke, Cooper, Cooray, Costantin, Cox, Croton, Daddi, Davé, de~la Vega,
  Dekel, Elbaz, Estrada-Carpenter, Faber, Fernández, Finkelstein, Freundlich,
  Fujimoto, Ángela García-Argumánez, Gardner, Gawiser, Gómez-Guijarro, Guo,
  Hamblin, Hamilton, Hathi, Holwerda, Hirschmann, Hutchison, Jaskot, Jha,
  Jogee, Juneau, Jung, Kassin, Bail, Leung, Lucas, Magnelli, Mantha, Matharu,
  McGrath, McIntosh, Merlin, Mobasher, Newman, Nicholls, Pandya, Rafelski,
  Ronayne, Santini, Seillé, Shah, Shen, Simons, Snyder, Stanway, Straughn,
  Teplitz, Vanderhoof, Vega-Ferrero, Wang, Weiner, Willmer, Wuyts, \&
  Zavala}]{Finkelstein2022_maisie}
Finkelstein, S.~L., Bagley, M.~B., Haro, P.~A., {et~al.} 2022, The
  Astrophysical Journal Letters, 940, L55

\bibitem[{{Finkelstein} {et~al.}(2023{\natexlab{b}}){Finkelstein}, {Leung},
  {Bagley}, {Dickinson}, {Ferguson}, {Papovich}, {Akins}, {Arrabal Haro},
  {Dave}, {Dekel}, {Kartaltepe}, {Kocevski}, {Koekemoer}, {Pirzkal},
  {Somerville}, {Yung}, {Amorin}, {Backhaus}, {Behroozi}, {Bisigello}, {Bromm},
  {Casey}, {Chavez Ortiz}, {Cheng}, {Chworowsky}, {Cleri}, {Cooper}, {Davis},
  {de la Vega}, {Elbaz}, {Franco}, {Fontana}, {Fujimoto}, {Giavalisco},
  {Grogin}, {Holwerda}, {Huertas-Company}, {Hirschmann}, {Iyer}, {Jogee},
  {Jung}, {Larson}, {Lucas}, {Mobasher}, {Morales}, {Morley}, {Mukherjee},
  {Perez-Gonzalez}, {Ravindranath}, {Rodighiero}, {Rowland}, {Tacchella},
  {Taylor}, {Trump}, \& {Wilkins}}]{Finkelstein2023a}
{Finkelstein}, S.~L., {Leung}, G. C.~K., {Bagley}, M.~B., {et~al.}
  2023{\natexlab{b}}, arXiv e-prints, arXiv:2311.04279

\bibitem[{Fudamoto {et~al.}(2020)Fudamoto, Oesch, Faisst, Béthermin, Ginolfi,
  Khusanova, Loiacono, Fèvre, Capak, Schaerer, Silverman, Cassata, Yan,
  Amorin, Bardelli, Boquien, Cimatti, Dessauges-Zavadsky, Fujimoto, Gruppioni,
  Hathi, Ibar, Jones, Koekemoer, Lagache, Lemaux, Maiolino, Narayanan, Pozzi,
  Riechers, Rodighiero, Talia, Toft, Vallini, Vergani, Zamorani, \&
  Zucca}]{Fudamoto2020}
Fudamoto, Y., Oesch, P.~A., Faisst, A., {et~al.} 2020, Astronomy \&
  Astrophysics, 643, A4

\bibitem[{{Fujimoto} {et~al.}(2023){Fujimoto}, {Finkelstein}, {Burgarella},
  {Carilli}, {Buat}, {Casey}, {Ciesla}, {Tacchella}, {Zavala}, {Brammer},
  {Fudamoto}, {Ouchi}, {Valentino}, {Cooper}, {Dickinson}, {Franco},
  {Giavalisco}, {Hutchison}, {Kartaltepe}, {Koekemoer}, {Kojima}, {Larson},
  {Murphy}, {Papovich}, {P{\'e}rez-Gonz{\'a}lez}, {Somerville}, {Yoon},
  {Wilkins}, {Akins}, {Amor{\'\i}n}, {Arrabal Haro}, {Bagley}, {Chworowsky},
  {Cleri}, {Cooper}, {Costantin}, {Daddi}, {Ferguson}, {Grogin},
  {Jim{\'e}nez-Andrade}, {Juneau}, {Kirkpatrick}, {Kocevski}, {Le Bail},
  {Long}, {Lucas}, {Magnelli}, {McKinney}, {Rose}, {Seill{\'e}}, {Simons},
  {Weiner}, \& {Yung}}]{Fujimoto2023}
{Fujimoto}, S., {Finkelstein}, S.~L., {Burgarella}, D., {et~al.} 2023, \apj,
  955, 130

\bibitem[{{Harikane} {et~al.}(2023){Harikane}, {Nakajima}, {Ouchi}, {Umeda},
  {Isobe}, {Ono}, {Xu}, \& {Zhang}}]{Harikane2023_specUVLF}
{Harikane}, Y., {Nakajima}, K., {Ouchi}, M., {et~al.} 2023, arXiv e-prints,
  arXiv:2304.06658

\bibitem[{Harikane {et~al.}(2023)Harikane, Ouchi, Oguri, Ono, Nakajima, Isobe,
  Umeda, Mawatari, \& Zhang}]{Harikane2023}
Harikane, Y., Ouchi, M., Oguri, M., {et~al.} 2023, The Astrophysical Journal
  Supplement Series, 265, 5

\bibitem[{Harris {et~al.}(2020)Harris, Millman, van~der Walt, Gommers,
  Virtanen, Cournapeau, Wieser, Taylor, Berg, Smith, Kern, Picus, Hoyer, van
  Kerkwijk, Brett, Haldane, del Río, Wiebe, Peterson, Gérard-Marchant,
  Sheppard, Reddy, Weckesser, Abbasi, Gohlke, \& Oliphant}]{Numpy2020}
Harris, C.~R., Millman, K.~J., van~der Walt, S.~J., {et~al.} 2020, Nature, 585,
  357

\bibitem[{Hunter(2007)}]{Hunter2007}
Hunter, J.~D. 2007, Computing in Science and Engineering, 9, 90

\bibitem[{Johnson {et~al.}(2021)Johnson, Leja, Conroy, \&
  Speagle}]{Johnson2021}
Johnson, B.~D., Leja, J., Conroy, C., \& Speagle, J.~S. 2021, The Astrophysical
  Journal Supplement Series, 254, 22

\bibitem[{{Kaasinen} {et~al.}(2023){Kaasinen}, {van Marrewijk}, {Popping},
  {Ginolfi}, {Di Mascolo}, {Mroczkowski}, {Concas}, {Di Cesare}, {Killi}, \&
  {Langan}}]{Kaasinen2023}
{Kaasinen}, M., {van Marrewijk}, J., {Popping}, G., {et~al.} 2023, \aap, 671,
  A29

\bibitem[{Kennicutt \& Evans(2012)}]{Kennicutt2012}
Kennicutt, R.~C. \& Evans, N.~J. 2012, Annual Review of Astronomy and
  Astrophysics, 50, 531

\bibitem[{{Kroupa}(2001)}]{Kroupa2001}
{Kroupa}, P. 2001, \mnras, 322, 231

\bibitem[{{Larson} {et~al.}(2023){Larson}, {Hutchison}, {Bagley},
  {Finkelstein}, {Yung}, {Somerville}, {Hirschmann}, {Brammer}, {Holwerda},
  {Papovich}, {Morales}, \& {Wilkins}}]{Larson2022b}
{Larson}, R.~L., {Hutchison}, T.~A., {Bagley}, M., {et~al.} 2023, \apj, 958,
  141

\bibitem[{Madau \& Dickinson(2014)}]{Madau2014}
Madau, P. \& Dickinson, M. 2014, Annual Review of Astronomy and Astrophysics,
  52, 415

\bibitem[{{Naidu} {et~al.}(2022){Naidu}, {Oesch}, {Setton}, {Matthee},
  {Conroy}, {Johnson}, {Weaver}, {Bouwens}, {Brammer}, {Dayal}, {Illingworth},
  {Barrufet}, {Belli}, {Bezanson}, {Bose}, {Heintz}, {Leja}, {Leonova},
  {Marques-Chaves}, {Stefanon}, {Toft}, {van der Wel}, {van Dokkum}, {Weibel},
  \& {Whitaker}}]{Naidu2022a}
{Naidu}, R.~P., {Oesch}, P.~A., {Setton}, D.~J., {et~al.} 2022, arXiv e-prints,
  arXiv:2208.02794

\bibitem[{Naidu {et~al.}(2022)Naidu, Oesch, van Dokkum, Nelson, Suess, Brammer,
  Whitaker, Illingworth, Bouwens, Tacchella, Matthee, Allen, Bezanson, Conroy,
  Labbe, Leja, Leonova, Magee, Price, Setton, Strait, Stefanon, Toft, Weaver,
  \& Weibel}]{Naidu2022}
Naidu, R.~P., Oesch, P.~A., van Dokkum, P., {et~al.} 2022, The Astrophysical
  Journal Letters, 940, L14

\bibitem[{{Nelson} {et~al.}(2023){Nelson}, {Suess}, {Bezanson}, {Price}, {van
  Dokkum}, {Leja}, {Wang}, {Whitaker}, {Labb{\'e}}, {Barrufet}, {Brammer},
  {Eisenstein}, {Gibson}, {Hartley}, {Johnson}, {Heintz}, {Mathews}, {Miller},
  {Oesch}, {Sandles}, {Setton}, {Speagle}, {Tacchella}, {Tadaki}, {{\"U}bler},
  \& {Weaver}}]{Nelson2022}
{Nelson}, E.~J., {Suess}, K.~A., {Bezanson}, R., {et~al.} 2023, \apjl, 948, L18

\bibitem[{Oke(1974)}]{Oke1974}
Oke, J.~B. 1974, The Astrophysical Journal Supplement Series, 27, 21

\bibitem[{{Popping}(2023)}]{Popping2023}
{Popping}, G. 2023, \aap, 669, L8

\bibitem[{Pérez-González {et~al.}(2023)Pérez-González, Barro,
  Annunziatella, Costantin, Ángela García-Argumánez, McGrath, Mérida,
  Zavala, Haro, Bagley, Backhaus, Behroozi, Bell, Bisigello, Buat, Calabrò,
  Casey, Cleri, Coogan, Cooper, Cooray, Dekel, Dickinson, Elbaz, Ferguson,
  Finkelstein, Fontana, Franco, Gardner, Giavalisco, Gómez-Guijarro, Grazian,
  Grogin, Guo, Huertas-Company, Jogee, Kartaltepe, Kewley, Kirkpatrick,
  Kocevski, Koekemoer, Long, Lotz, Lucas, Papovich, Pirzkal, Ravindranath,
  Somerville, Tacchella, Trump, Wang, Wilkins, Wuyts, Yang, \&
  Yung}]{Perez-Gonzalez2022}
Pérez-González, P.~G., Barro, G., Annunziatella, M., {et~al.} 2023, The
  Astrophysical Journal Letters, 946, L16

\bibitem[{Rieke {et~al.}(2023)Rieke, Kelly, Misselt, Stansberry, Boyer, Beatty,
  Egami, Florian, Greene, Hainline, Leisenring, Roellig, Schlawin, Sun, Tinnin,
  Williams, Willmer, Wilson, Clark, Rohrbach, Brooks, Canipe, Correnti,
  DiFelice, Gennaro, Girard, Hartig, Hilbert, Koekemoer, Nikolov, Pirzkal,
  Rest, Robberto, Sunnquist, Telfer, Wu, Ferry, Lewis, Baum, Beichman, Doyon,
  Dressler, Eisenstein, Ferrarese, Hodapp, Horner, Jaffe, Johnstone, Krist,
  Martin, McCarthy, Meyer, Rieke, Trauger, \& Young}]{nircam_performance}
Rieke, M.~J., Kelly, D.~M., Misselt, K., {et~al.} 2023, Publications of the
  Astronomical Society of the Pacific, 135, 028001

\bibitem[{Rodighiero {et~al.}(2022)Rodighiero, Bisigello, Iani, Marasco,
  Grazian, Sinigaglia, Cassata, \& Gruppioni}]{Rodighiero2022}
Rodighiero, G., Bisigello, L., Iani, E., {et~al.} 2022, Monthly Notices of the
  Royal Astronomical Society: Letters, 518, L19

\bibitem[{Schreiber {et~al.}(2015)Schreiber, Pannella, Elbaz, Béthermin,
  Inami, Dickinson, Magnelli, Wang, Aussel, Daddi, Juneau, Shu, Sargent, Buat,
  Faber, Ferguson, Giavalisco, Koekemoer, Magdis, Morrison, Papovich, Santini,
  \& Scott}]{Schreiber2015}
Schreiber, C., Pannella, M., Elbaz, D., {et~al.} 2015, Astronomy \&
  Astrophysics, 575, A74

\bibitem[{{Shu} {et~al.}(2022){Shu}, {Yang}, {Liu}, {Wang}, {Wang}, {Han},
  {Huang}, {Lim}, {Chang}, {Zheng}, {Zheng}, {Wang}, \& {Kong}}]{Shu2022}
{Shu}, X., {Yang}, L., {Liu}, D., {et~al.} 2022, \apj, 926, 155

\bibitem[{{Smail} {et~al.}(2023){Smail}, {Dudzevi{\v{c}}i{\={u}}t{\.{e}}},
  {Gurwell}, {Fazio}, {Willner}, {Swinbank}, {Arumugam}, {Summers}, {Cohen},
  {Jansen}, {Windhorst}, {Meena}, {Zitrin}, {Keel}, {Cheng}, {Coe},
  {Conselice}, {D'Silva}, {Driver}, {Frye}, {Grogin}, {Koekemoer}, {Marshall},
  {Nonino}, {Pirzkal}, {Robotham}, {Rutkowski}, {Ryan}, {Tompkins}, {Willmer},
  {Yan}, {Broadhurst}, {Diego}, {Kamieneski}, \& {Yun}}]{Smail2023}
{Smail}, I., {Dudzevi{\v{c}}i{\={u}}t{\.{e}}}, U., {Gurwell}, M., {et~al.}
  2023, \apj, 958, 36

\bibitem[{{Smail} {et~al.}(2002){Smail}, {Ivison}, {Blain}, \&
  {Kneib}}]{SMail2002}
{Smail}, I., {Ivison}, R.~J., {Blain}, A.~W., \& {Kneib}, J.~P. 2002, \mnras,
  331, 495

\bibitem[{Stalevski {et~al.}(2016)Stalevski, Ricci, Ueda, Lira, Fritz, \&
  Baes}]{Stalevski2016}
Stalevski, M., Ricci, C., Ueda, Y., {et~al.} 2016, Monthly Notices of the Royal
  Astronomical Society, 458, 2288

\bibitem[{Swinbank {et~al.}(2014)Swinbank, Simpson, Smail, Harrison, Hodge,
  Karim, Walter, Alexander, Brandt, de~Breuck, da~Cunha, Chapman, Coppin,
  Danielson, Dannerbauer, Decarli, Greve, Ivison, Knudsen, Lagos, Schinnerer,
  Thomson, Wardlow, Weiß, \& van~der Werf}]{Swinbank2014}
Swinbank, A.~M., Simpson, J.~M., Smail, I., {et~al.} 2014, Monthly Notices of
  the Royal Astronomical Society, 438, 1267

\bibitem[{Tacchella {et~al.}(2022)Tacchella, Johnson, Robertson, Carniani,
  D'Eugenio, Kumar, Maiolino, Nelson, Suess, Übler, Williams, Adebusola,
  Alberts, Arribas, Bhatawdekar, Bonaventura, Bowler, Bunker, Cameron, Curti,
  Egami, Eisenstein, Frye, Hainline, Helton, Ji, Looser, Lyu, Perna, Rawle,
  Rieke, Rieke, Saxena, Sandles, Shivaei, Simmonds, Sun, Willmer, Willott, \&
  Witstok}]{Tacchella2022}
Tacchella, S., Johnson, B.~D., Robertson, B.~E., {et~al.} 2022, MNRAS, 000, 1

\bibitem[{Virtanen {et~al.}(2020)Virtanen, Gommers, Oliphant, Haberland, Reddy,
  Cournapeau, Burovski, Peterson, Weckesser, Bright, van~der Walt, Brett,
  Wilson, Millman, Mayorov, Nelson, Jones, Kern, Larson, Carey, İlhan Polat,
  Feng, Moore, VanderPlas, Laxalde, Perktold, Cimrman, Henriksen, Quintero,
  Harris, Archibald, Ribeiro, Pedregosa, van Mulbregt, Vijaykumar, Bardelli,
  Rothberg, Hilboll, Kloeckner, Scopatz, Lee, Rokem, Woods, Fulton, Masson,
  Häggström, Fitzgerald, Nicholson, Hagen, Pasechnik, Olivetti, Martin,
  Wieser, Silva, Lenders, Wilhelm, Young, Price, Ingold, Allen, Lee, Audren,
  Probst, Dietrich, Silterra, Webber, Slavič, Nothman, Buchner, Kulick,
  Schönberger, de~Miranda~Cardoso, Reimer, Harrington, Rodríguez,
  Nunez-Iglesias, Kuczynski, Tritz, Thoma, Newville, Kümmerer, Bolingbroke,
  Tartre, Pak, Smith, Nowaczyk, Shebanov, Pavlyk, Brodtkorb, Lee, McGibbon,
  Feldbauer, Lewis, Tygier, Sievert, Vigna, Peterson, More, Pudlik, Oshima,
  Pingel, Robitaille, Spura, Jones, Cera, Leslie, Zito, Krauss, Upadhyay,
  Halchenko, Vázquez-Baeza, \& Contributors}]{Virtanen2020}
Virtanen, P., Gommers, R., Oliphant, T.~E., {et~al.} 2020, Nature Methods, 17,
  261

\bibitem[{Wang {et~al.}(2023)Wang, Leja, Bezanson, Johnson, Khullar, Labbé,
  Price, Weaver, \& Whitaker}]{WangB2023}
Wang, B., Leja, J., Bezanson, R., {et~al.} 2023, The Astrophysical Journal
  Letters, 944, L58

\bibitem[{Wang {et~al.}(2019)Wang, Schreiber, Elbaz, Yoshimura, Kohno, Shu,
  Yamaguchi, Pannella, Franco, Huang, Lim, \& Wang}]{TWang2019}
Wang, T., Schreiber, C., Elbaz, D., {et~al.} 2019, Nature, 572, 211

\bibitem[{Williams {et~al.}(2019)Williams, Labbe, Spilker, Stefanon, Leja,
  Whitaker, Bezanson, Narayanan, Oesch, Weiner, Williams, Labbe, Spilker,
  Stefanon, Leja, Whitaker, Bezanson, Narayanan, Oesch, \&
  Weiner}]{Williams2019}
Williams, C.~C., Labbe, I., Spilker, J., {et~al.} 2019, ApJ, 884, 154

\bibitem[{Windhorst {et~al.}(2023)Windhorst, Cohen, Jansen, Summers, Tompkins,
  Conselice, Driver, Yan, Coe, Frye, Grogin, Koekemoer, Marshall, O’Brien,
  Pirzkal, Robotham, Ryan, Willmer, Carleton, Diego, Keel, Porto, Redshaw,
  Scheller, Wilkins, Willner, Zitrin, Adams, Austin, Arendt, Beacom,
  Bhatawdekar, Bradley, Broadhurst, Cheng, Civano, Dai, Dole, D’Silva,
  Duncan, Fazio, Ferrami, Ferreira, Finkelstein, Furtak, Gim, Griffiths,
  Hammel, Harrington, Hathi, Holwerda, Honor, Huang, Hyun, Im, Joshi,
  Kamieneski, Kelly, Larson, Li, Lim, Ma, Maksym, Manzoni, Meena, Milam,
  Nonino, Pascale, Petric, Pierel, del Carmen~Polletta, Röttgering, Rutkowski,
  Smail, Straughn, Strolger, Swirbul, Trussler, Wang, Welch, Wyithe, Yun,
  Zackrisson, Zhang, \& Zhao}]{Windhorst2023}
Windhorst, R.~A., Cohen, S.~H., Jansen, R.~A., {et~al.} 2023, The Astronomical
  Journal, 165, 13

\bibitem[{Xiao {et~al.}(2023)Xiao, Elbaz, Gómez-Guijarro, Leroy, Bing, Daddi,
  Magnelli, Franco, Zhou, Dickinson, Wang, Rujopakarn, Magdis, Treister, Inami,
  Demarco, Sargent, Shu, Kartaltepe, Alexander, Béthermin, Bournaud, Ciesla,
  Ferguson, Finkelstein, Giavalisco, Gu, Iono, Juneau, Lagache, Leiton,
  Messias, Motohara, Mullaney, Nagar, Pannella, Papovich, Pope, Schreiber, \&
  Silverman}]{Xiao2023}
Xiao, M.-Y., Elbaz, D., Gómez-Guijarro, C., {et~al.} 2023, Astronomy \&
  Astrophysics, 672, A18

\bibitem[{Yan {et~al.}(2023{\natexlab{a}})Yan, Cohen, Windhorst, Jansen, Ma,
  Beacom, Ling, Cheng, Huang, Grogin, Willner, Yun, Hammel, Milam, Conselice,
  Driver, Frye, Marshall, Koekemoer, Willmer, Robotham, D’Silva, Summers,
  Lim, Harrington, Ferreira, Diego, Pirzkal, Wilkins, Wang, Hathi, Zitrin,
  Bhatawdekar, Adams, Furtak, Maksym, Rutkowski, \& Fazio}]{Yan2023_darkfield}
Yan, H., Cohen, S.~H., Windhorst, R.~A., {et~al.} 2023{\natexlab{a}}, The
  Astrophysical Journal Letters, 942, L8

\bibitem[{Yan {et~al.}(2023{\natexlab{b}})Yan, Ma, Ling, Cheng, \&
  Huang}]{Yan2023_ero}
Yan, H., Ma, Z., Ling, C., Cheng, C., \& Huang, J.-S. 2023{\natexlab{b}}, The
  Astrophysical Journal Letters, 942, L9

\bibitem[{Zavala {et~al.}(2023)Zavala, Buat, Casey, Finkelstein, Burgarella,
  Bagley, Ciesla, Daddi, Dickinson, Ferguson, Franco, Jiménez-Andrade,
  Kartaltepe, Koekemoer, Bail, Murphy, Papovich, Tacchella, Wilkins, Aretxaga,
  Behroozi, Champagne, Fontana, Giavalisco, Grazian, Grogin, Kewley, Kocevski,
  Kirkpatrick, Lotz, Pentericci, Pérez-González, Pirzkal, Ravindranath,
  Somerville, Trump, Yang, Yung, Almaini, Amorín, Annunziatella, Haro,
  Backhaus, Barro, Bell, Bhatawdekar, Bisigello, Buitrago, Calabrò,
  Castellano, Óscar A.~Chávez~Ortiz, Chworowsky, Cleri, Cohen, Cole, Cooke,
  Cooper, Cooray, Costantin, Cox, Croton, Davé, de~la Vega, Dekel, Elbaz,
  Estrada-Carpenter, Fernández, Finkelstein, Freundlich, Fujimoto, Ángela
  García-Argumánez, Gardner, Gawiser, Gómez-Guijarro, Guo, Hamilton, Hathi,
  Holwerda, Hirschmann, Huertas-Company, Hutchison, Iyer, Jaskot, Jha, Jogee,
  Juneau, Jung, Kassin, Kurczynski, Larson, Leung, Long, Lucas, Magnelli,
  Mantha, Matharu, McGrath, McIntosh, Medrano, Merlin, Mobasher, Morales,
  Newman, Nicholls, Pandya, Rafelski, Ronayne, Rose, Ryan, Santini, Seillé,
  Shah, Shen, Simons, Snyder, Stanway, Straughn, Teplitz, Vanderhoof,
  Vega-Ferrero, Wang, Weiner, Willmer, \& Wuyts}]{Zavala2023_z5dusty}
Zavala, J.~A., Buat, V., Casey, C.~M., {et~al.} 2023, The Astrophysical Journal
  Letters, 943, L9

\end{thebibliography}

\appendix

\section{Obscured SFR and rest-frame FIR SED predictions from JWST-only photometry}
\label{appendix:A}
In this appendix, we explore the low-, intermediate-, and high-redshift solutions of \emph{BAGPIPES}, \emph{CIGALE} and \emph{Prospector} fitted only to the HST and JWST photometry of the red objects considered in this work. In particular, we show in Fig. 1, the predictions for the rest-frame FIR continuum fluxes and the SFR implied by the red JWST slopes as a function of redshift. For \emph{CIGALE} and \emph{Prospector} we show only the highest likelihood solutions, and for \emph{BAGPIPES} we sampled the full posterior as it is quite compact in parameter space. We find that for all SED codes, the inferred SFR and observed sub-mm fluxes increase strongly as a function of redshift. In fact, the inferred SFR are in tension with the \textit{extrapolated} CSFRD at $z>8$. This suggests that CSFRD priors could be used for large samples of JWST objects in wide fields to refine photometric redshifts. For \emph{BAGPIPES} and \emph{CIGALE}, the NOEMA fluxes are predicted to rise by two or three orders of magnitudes, between $z=2$ and $z>10$, with relatively small scatter at any redshift (factor $2-3\times$). The NOEMA flux ratio between the two sidebands of a single Polyfix tuning ($\Delta \nu_{\rm{obs}} = 15.488\rm{GHz}$) evolves quasi-linearly with redshift, reflecting the simple dust continuum emission model. The situation is more nuanced with \emph{Prospector}, where more complex dust modelling leads to a larger scatter in the SFR and predicted dust continuum emission, albeit with the same redshift trends. Therefore, we conclude that ALMA/NOEMA should identify red $z>10$ candidates with $m_{AB}(F444W)\sim 25$ in $\sim 5-30$ min. For objects with blue rest-frame UV colours, the situation is different as the presence of strong continuum dust emission is not expected \citep[e.g.][]{Kaasinen2023,Bakx2023, Popping2023,Fujimoto2023}.
\begin{figure*}[h!]
    \centering
    \includegraphics[width=0.8\textwidth]{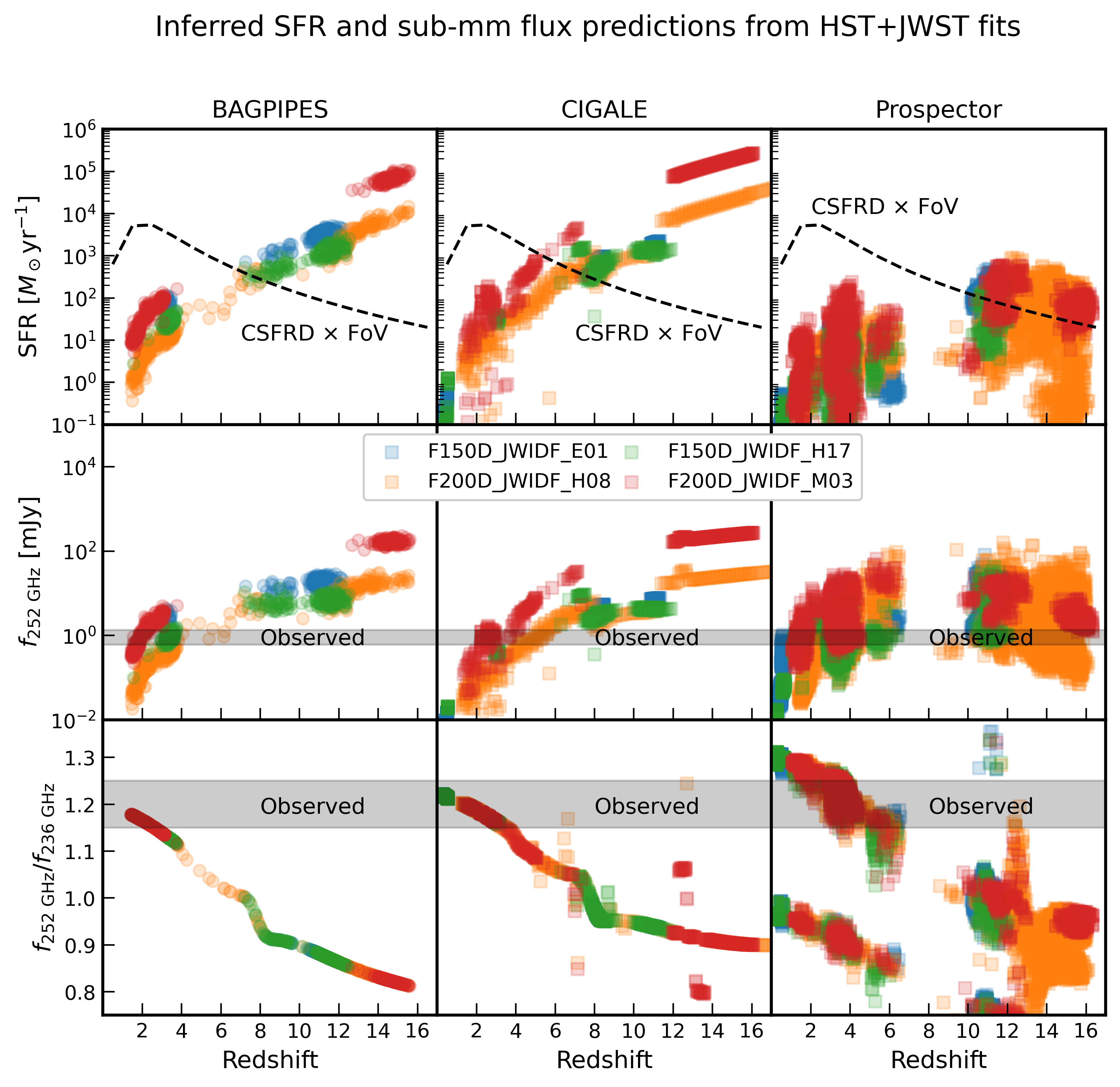}
    \caption{Inferred SFR and sub-mm flux predictions from JWST and HST using three different SED codes. We show the inferred SFR in the first row, and sub-mm flux predictions in second and third rows. The three columns show, in order, the predictions for \emph{BAGPIPES}, \emph{CIGALE} and \emph{Prospector} fits to the HST and JWST photometry of the red dropouts considered in this work (coloured points). For \emph{CIGALE} and \emph{Prospector,} we show only the highest likelihood solutions $\Delta L \sim 1$; while for \emph{BAGPIPES,} we sample the full posterior as it is quite compact in parameter space. In the first row, we overplot the CSFRD multiplied by the field of view of the IRAC Dark Field (IDF), e.g. giving the SFR of a single galaxy contributing $100\%$ to the CSFRD in that field. In the second and third row we overlay the different predictions with a horizontal gray bar the observed NOEMA fluxes and flux ratio. By comparing the observed NOEMA fluxes and flux ratios, we show clearly that the lower redshift solutions are strongly preferred. In the lower panels, the main trend in the NOEMA flux ratio is due to the dust graybody emission profile redshifting across the Polyfix sidebands. The outliers (\emph{CIGALE}) and second track (\emph{Prospector}) are due to high dust temperatures absent from the \emph{BAGPIPES} fits. }
    \label{fig:predictions_redshift}
\end{figure*}

\end{document}